\documentclass[aps,pre,twocolumn,showpacs,amssymb,longbibliography]{revtex4-1}

\usepackage[centertags]{amsmath}
\usepackage{graphicx}
\usepackage{amsfonts}
\usepackage[normalem]{ulem} 
\usepackage{color}
\usepackage{psfrag}

\begin{document}

\graphicspath{{Figures_bold_review/}}

\title{Time-Reversal of Nonlinear Waves - Applicability and Limitations}
\author{G. Ducrozet$^{1*}$, M. Fink$^{2}$ and A. Chabchoub$^{3,4}$}
\affiliation{$^1$ \'Ecole Centrale Nantes, LHEEA Lab., UMR CNRS 6598, 1 rue de la No\"e, 44321 Nantes, France}
\email{guillaume.ducrozet@ec-nantes.fr}
\affiliation{$^2$ Institut Langevin, ESPCI \& CNRS, UMR CNRS 7587, 10 rue Vauquelin, 75005 Paris, France}
\affiliation{$^3$ Department of Mechanical Engineering, School of Engineering, Aalto University, 02150 Espoo, Finland}
\affiliation{$^4$ Department of Ocean Technology Policy and Environment, Graduate School of Frontier Sciences, The University of Tokyo, Kashiwa, Chiba 277-8563, Japan}

\begin{abstract}
Time-reversal (TR) refocusing of waves is one of fundamental principles in wave physics. Using the TR approach, ``Time-reversal mirrors'' can physically create a time-reversed wave that exactly refocus back, in space and time, to its original source regardless of the complexity of the medium as if time were going backwards. Lately, laboratory experiments proved that this approach can be applied not only in acoustics and electromagnetism but also in the field of linear and nonlinear water waves. Studying the range of validity and limitations of the TR approach may determine and quantify its range of applicability in hydrodynamics. In this context, we report a numerical study of hydrodynamic TR using a uni-directional numerical wave tank, implemented by the nonlinear high-order spectral method, known to accurately model the physical processes at play, beyond physical laboratory restrictions. The applicability of the TR approach is assessed over a variety of hydrodynamic localized and pulsating structures' configurations, pointing out the importance of high-order dispersive and particularly nonlinear effects in the refocusing of hydrodynamic stationary envelope solitons and breathers. We expect that the results may motivate similar experiments in other nonlinear dispersive media and encourage several applications with particular emphasis on the field of ocean engineering. 
\end{abstract}

\maketitle

\section{Introduction}

Nonlinearity is a fundamental feature of hydrodynamic evolution equations and is therefore crucial for the accurate description of water waves \cite{stokes1847theory,mei1983applied}. One particular and prominent form of hydrodynamic instability of nonlinear waves is the Benjamin-Feir instability \cite{benjamin1967disintegration}. This instability has been discussed in modeling oceanic extremes. Indeed, the formation of extremely large and steep waves in real seas, also known as freak or rogue waves (RWs), has drawn significant scientific attention in past years, see \cite{chabchoub2011rogue,kibler2010peregrine,solli2007optical,onorato2013rogue,dudley2014instabilities}. In fact, the dynamics of RWs can be described and modeled within the framework of exact solutions of the nonlinear Schr\"odinger equation (NLS) \cite{zakharov1968stability,kharif2003physical,osborne2010nonlinear}. In fact, integrable weakly nonlinear evolution equations, such as the NLS, allow to discuss the propagation characteristics of specific and complex waves by means of exact solutions \cite{zabusky1965interaction,yuen1982nonlinear}. Recent laboratory as well as numerical studies confirmed the existence of hydrodynamic breathers \cite{chabchoub2012observation,chabchoub2014hydrodynamics,slunyaev2013super}. Being controllable in both time and space, breathers are more accurate physical models to generate extreme waves for engineering application purposes, rather than using the inefficient as well as simplified linear superposition principle. Prominent models are for instance the family of doubly-localized Peregrine and Akhmediev-Peregrine breathers.

Time-reversal (TR) procedure is another fundamental principle in wave physics and it has been used in several contexts in past years, especially, for acoustic and elastic waves \cite{fink1992time,roux2000time}. This TR technique, based on TR mirrors works as the following: a given wave pulse, generated at a source, is then measured by a set of antennas or probes after a certain propagation in space. These signals are then time-reversed and rebroadcasted from the same measurement locations by the TR mirrors. If TR-invariance is valid, the initial wave pulse is expected to focus back at the original source location, independently of the complexity of the medium.

The effects of dispersion \cite{roux2000time,ros1998time} and nonlinearities \cite{tanter2001breaking} have been experimentally studied for acoustic waves. It has been shown that the time-reversed field focuses back in time and space as long as nonlinearities do not create dissipation, i.e., as long as the propagation distance is smaller than the shock distance.

Different TR procedures have been experimentally applied in the field of water waves. The TR mirror has been shown to be valid for wave fields with very weak nonlinearity in the configuration of a water tank cavity and multiple mirror reflections \cite{przadka2012time} and recently, a new concept of instantaneous time mirror has been also tested for water waves \cite{bacot2016time}. As next, strongly and doubly-localized breather solutions have been used to study the time refocusing of nonlinear wave in a uni-directional wave flume \cite{chabchoub2014time}. This study has shown that for these RW models which are steep in amplitude and therefore strongly nonlinear waves, TR is still accurately valid. 

In this paper, we extend the experimental work \cite{chabchoub2014time} and report an extensive and detailed numerical study of the effects of dispersion and nonlinearities on hydrodynamic TR within the framework of NLS envelope solitons and breathers. The main goal of this work is to accurately quantify the validity of the TR approach for water waves, based on a numerical wave tank (NWT) \cite{ducrozet2012modified}, implemented by the conservative higher-order spectral method (HOS) \cite{west1987new,dommermuth1987high}. It is shown that the TR is indeed valid for nonlinear waves for a wide range of wave propagation distances as well as steepness values. This proves the robustness of TR property of NLS localized structures with respect to noise and chaos. Limitations of the TR approach are also discussed in detail. We emphasize that the investigation of the effect of dispersion and nonlinearity on hydrodynamic time-reversal characteristics of focused waves, taking also into account for instance bound waves and the inaccuracies of wave maker's wave generation has been studied within the framework of Zakharov equation in \cite{shemer2007evolution}.

The paper is organized as follows: the first section presents the weakly nonlinear theory of water waves and the TR principle. Then, the NWT configuration will be detailed as well as the method used to generate localized structures inside this NWT. The third section is devoted to the validity analysis of the TR for water waves in relation to nonlinearity of the wave field as well as to dispersion effects. Finally, the last section analyzes the properties of the physical processes, including a discussion on the limitations of the TR method in hydrodynamics.

\section{Nonlinear waves and the Time-Reversal principle}
\label{sec:NonlinearWaves_TR}

In this work, we are interested in the propagation of water waves at large scales and consequently we limit our study to surface gravity waves. Furthermore, we assume that the waves are evolving in deep-water, that is, we assume that the ratio of water depth $h$ to the wavelength $\lambda$ is large. For practical applications, a deep-water condition is already satisfied when $h/\lambda > 1/2$. Finally, we are interested in configurations in which dissipation is negligible and consequently do not consider viscous effects of any kind: solid boundaries, wave breaking, etc.

This classical set of assumptions for water waves leads to the modeling of this physical phenomena thanks to the nonlinear potential flow theory. This has been widely used to study several nonlinear wave phenomena such as nonlinear wave interactions \cite{zakharov1968stability}, RWs \cite{ducrozet2007HOS,slunyaev2013highest} and water-wave turbulence \cite{nazarenko2011wave}. Numerical solutions of the fully nonlinear problem can be complex and very time-consuming from a numerical point of view. An alternative to overcome these constraints is to assume weak nonlinearity of the wave field, while processes are expected to be narrow-banded, allowing slow modulations. With those additional hypotheses, one may use the focusing NLS equation \cite{zakharov1968stability} to describe the evolution of nonlinear narrow-banded wave trains in deep-water

\begin{equation} \label{eq:NLS}
i \left( \dfrac{\partial \psi}{\partial t} + \dfrac{\omega}{2k} \dfrac{\partial \psi}{\partial x}\right) - \dfrac{\omega}{8k^2} \dfrac{\partial^2 \psi}{\partial x^2} - \dfrac{\omega k^2}{2} \left| \psi \right|^2 \psi=0, 
\end{equation}

\noindent
while $\psi(x,t)$ is the complex envelope of the corresponding wave train. To second-order of approximation in wave steepness $ka$ ($k=2\pi/\lambda$ is the wave number and $a$ stands for the wave amplitude), the free surface elevation is then given by

\begin{eqnarray} \label{eq:reconstruction_FS}
	\eta(x,t) &=& \operatorname{Re}\left( \psi(x,t) \exp\left( i \left[ kx-\omega t \right] \right) \right) \nonumber \\
	&& + \operatorname{Re}\left( \frac{k \psi^2(x,t)}{2} \exp\left( 2i \left[ kx-\omega t \right] \right) \right)
\end{eqnarray}

In the following, we consider exact analytic solutions to NLS to assess the validity of TR procedure. The simplest form is the stationary envelope solution \cite{shabat1972exact}. Most prominent form of breathers is the family of doubly-localized breathers, also referred to as Akhmediev-Peregrine breathers \cite{peregrine1983water,akhmediev1985generation}, that have been emphasized to be appropriate models to describe oceanic RWs \cite{shrira2010makes,akhmediev2009waves}. Detailed description and parameterization of those analytic solutions is given in Supplemental Material \cite{supplement}.\\

For a given solution $\psi(x,t)$ of \eqref{eq:NLS}, it can be easily verified that the time-reversed complex conjugate form $\psi^*(x,-t)$ is also a solution. Therefore, both corresponding free surface elevations $\eta(x,t)$ and $\eta(x,-t)$ are possible solutions to the water wave problem within the framework of the NLS. Thanks to this property, a TR mirror can be used to create the time-reversed hydrodynamic wave field $\eta(x,−t)$ in the whole propagating medium. 

The theory indicates that to be effective, the TR procedure needs to measure and time-reverse the free surface elevation as well as the normal velocity of this surface. However, from a practical point of view in water waves experiments, no simple generation procedure exists to ensure both those quantities at the same time. We consequently restrict our study to practical applications in laboratories (for instance ocean engineering facilities) in which the only free surface elevation is imposed/controlled by a wave maker. Indeed, it has been shown experimentally \cite{przadka2012time,chabchoub2014time} that this procedure is efficient. If the problem has only one horizontal spatial dimension, it appears sufficient to measure the wave field at one location $\eta(x_M,t)$ and rebroadcast the time-reserved signal $\eta(x_M,-t)$ from this unique point $x_M$ (mirror position) in order to observe the solution $\eta(x,-t)$ in the whole medium and thus, to verify the TR invariance.\\

Within a water wave basin, the possible procedure to verify the validity of TR procedure is: i) generate a given wave field, ii) measure the spread waves at a specific location, iii) time-reverse this signal and use it to generate a new wave field. If time-reversibility is valid, the wave field measured by the probe, during the second experiment, should correspond to pulse, generated by the wave maker's motion of the first experiment.

TR has been experimentally studied in a wave basin within the framework of breathers in \cite{chabchoub2014time} for initial solutions that exhibit strongly nonlinear features. In fact, doubly-localized breathers up to second-order have been accurately refocused through TR. This indicates that the TR procedure, in the presence of high nonlinearity and dispersive effects, is still valid. It allows the use of the time-reversible feature of the wave field, even in the case of strong focusing.\\

The purpose of the proposed study is to detail the range of applicability of the TR method for water waves in the configuration of a uni-directional wave basin. As stated previously, the time reversibility can be demonstrated in the context of the NLS equation and is also theoretically valid for the fully nonlinear conservative water wave problem (potential flow formalism). However, the practical TR procedure relies on some assumptions that may result in a loss of accurate wave refocusing (use of the only free surface elevation, conversion to the wave maker's motion, etc.). We will show that the application of TR procedure to NLS solutions is indeed valid for a significant range of parameters. Therefore, even if the target wave field is only an approximation to the fully nonlinear physics of the water wave problem (assuming weak nonlinearity and narrow-banded dynamics), TR procedure is applicable. In this concern, the important parameters for this study are expected to be: i) nonlinearity, characterized by the steepness of the wave field $ka$ and ii) higher-order dispersive terms, appearing to be non-negligible, when the process is broad-banded. The latter effects related to higher-order dispersion can be characterized through the propagating distance $kx_M$.

\section{Numerical wave tank configuration and generation of hydrodynamic localized structures}

Following the experiments presented in \cite{chabchoub2014time}, a numerical procedure is set to provide an extensive analysis with respect to the applicability and limitations, related to the use of the TR approach for nonlinear waves in a wave basin. This section presents the numerical model as well as the setup validation.

\subsection{Numerical Wave Tank}

The numerical validation of the TR procedure in the context of water waves relies on an accurate description of the wave physics at play during the wave propagation. Primarily, it has been discussed in previous section that possible limitations are due to high-order nonlinearities as well as high-order dispersive terms. Consequently, the suitable numerical model must include these physical features in the configuration of the nonlinear potential flow solver, being the typical candidate, if we exclude the breaking of the waves.\\

The HOS model \cite{west1987new,dommermuth1987high}, described briefly in Supplemental Material \cite{supplement}, has been widely used in the context of water waves for the study of specific nonlinear processes such as modulation instability \cite{toffoli2010evolution,slunyaev2013super}, simulation of RWs in irregular sea states \cite{ducrozet2007HOS} or emergence of bimodal seas \cite{toffoli2010development}. This highly nonlinear scheme allows an efficient and accurate solution of the water wave problem.

This original HOS model is however limited to the study of nonlinear waves propagating in a periodic domain, specifying an initial condition. The configuration is essentially different in the context of a wave basin. This exhibits a wave maker to generate a given wave field, an absorbing beach on the other side of the basin to prevent wave reflections as well as reflection conditions on the side walls. Consequently, specific attention has been paid to the development of a NWT, we refer to as HOS-NWT, that includes all those specificities. This model has been validated with several comparisons to experiments on different wave fields: from regular uni-directional waves to directional irregular sea states, see \cite{ducrozet2012modified} for details.

\begin{figure}[h!tbp]
\begin{center}
\includegraphics[width=\linewidth]{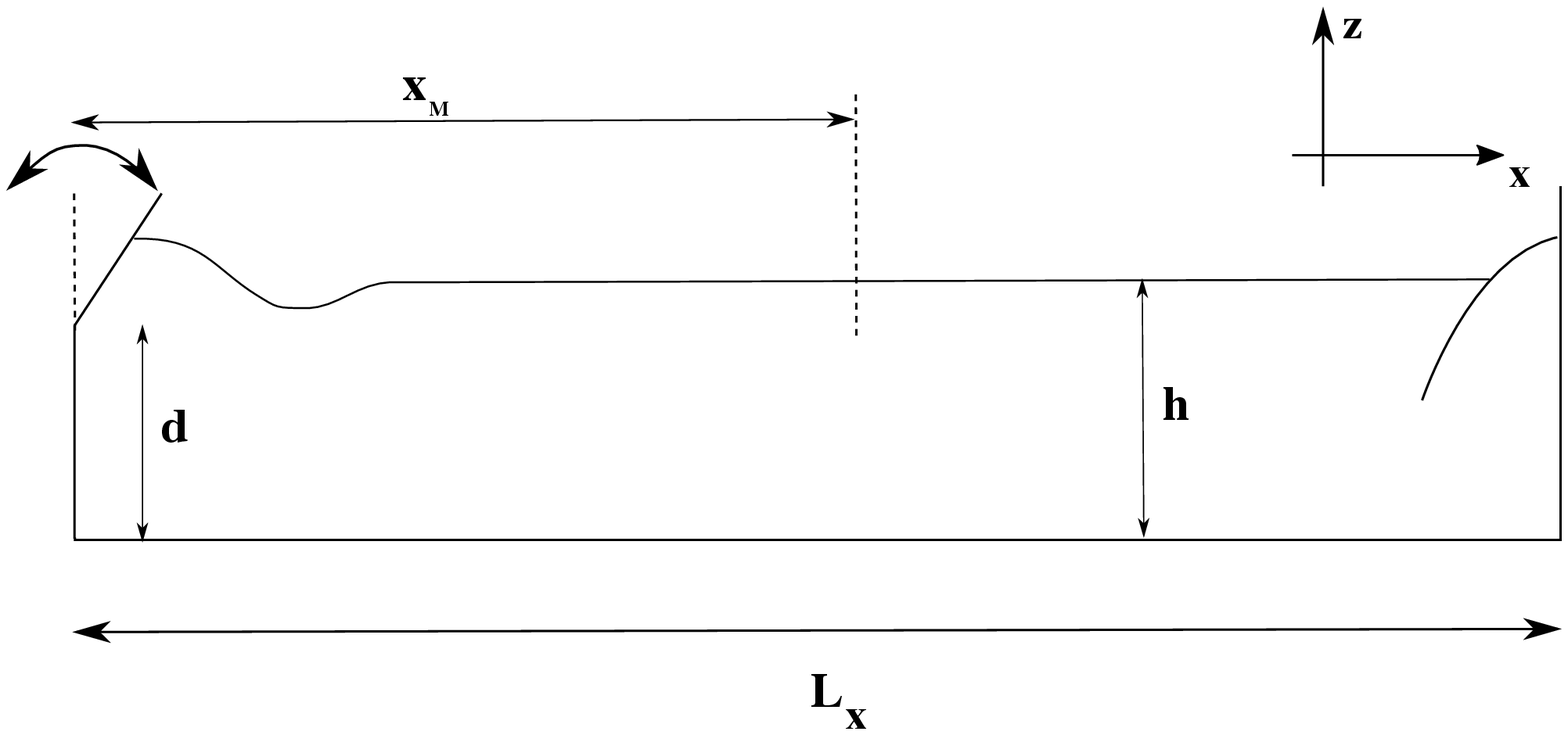}
\caption{Sketch of the NWT.}
\label{fig:NWT}
\end{center}
\end{figure}

In the following, the configuration used for the NWT is similar to the experiments reported in \cite{chabchoub2014time}. The uni-directional wave field will be analyzed in the configuration, sketched on Figure \ref{fig:NWT} for the sake of appropriate analysis dimensions. The wave maker is defined as a hinged flap and the numerical absorbing beach is designed so that no reflection occurs during the simulations. The water depth is defined as $h=1$ m, the hinged location is $d=0.8$ m, while the length of the basin is set to be $L_x=20$ m.

The numerical parameters in HOS-NWT have been chosen to ensure an accurate description of the physical phenomena at stake. These are detailed in Supplemental Material \cite{supplement}.

\subsection{Generation of a localized wave field}

We start the TR procedure by first defining the temporal surface wave profile of the pulse to be studied $\eta(x_S,t)$ at the source (wavemaker) position $x_S$. As specified earlier, different specific NLS pulses for several wave parameters will be investigated in this work. The choice of the analytic envelope soliton, or doubly-localized structures, will define the complex envelope $\psi(x,t)$ and the corresponding target free surface elevation at the maximal envelope amplitude $\eta(x=0,t)$ through Eq. \eqref{eq:reconstruction_FS}. This allows to define the wave maker's motion through the use of an adequate transfer function, dependent of the wave maker's shape. The TR procedure consequently relies on a linear approximation to relate the target free surface elevation to this motion, leading to possible discrepancies.

\begin{figure}[h!tbp]
\begin{center}
\includegraphics[width=\linewidth]{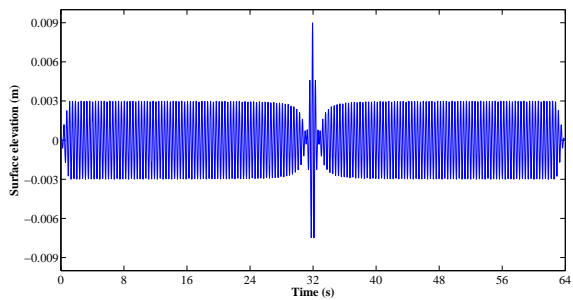}
\caption{Target wave elevation used to define the wave maker's motion: Peregrine breather with $ka=0.09$ and $a=0.003$ m, evaluated at $x=0$.}
\label{fig:Initial_Peregrine_ka_0.09}
\end{center}
\end{figure}

We demonstrate the procedure by means of the Peregrine solution. Figure \ref{fig:Initial_Peregrine_ka_0.09} gives an example of the free surface profile $\eta(x=0,t)$ used to deduce the wave maker's motion. It is the case of a Peregrine breather with carrier wave steepness $ka=0.09$ and a carrier amplitude $a=0.003$ m, evaluated at $x=0$. We used ramps at the beginning and end of the chosen time window in order to ensure: i) a smooth start of the wave maker's movement and ii) the periodicity of the signal, which is decomposed on Fourier components.

In the different configurations tested, the time-window of the target free surface elevation is kept constant and is specified and depicted on Figure \ref{fig:Initial_Peregrine_ka_0.09}. This corresponds to $T_w \simeq 175 T_0$, with $T_0=2\pi / \omega$ the carrier wave period. This is sufficiently long to prevent from any boundary effects on a sufficiently long time window around the modulation, that occurs around $\simeq 110~T_0$, as presented in \cite{chabchoub2014time}.\\

\begin{figure}[h!tbp]
\begin{center}
\includegraphics[width=\linewidth]{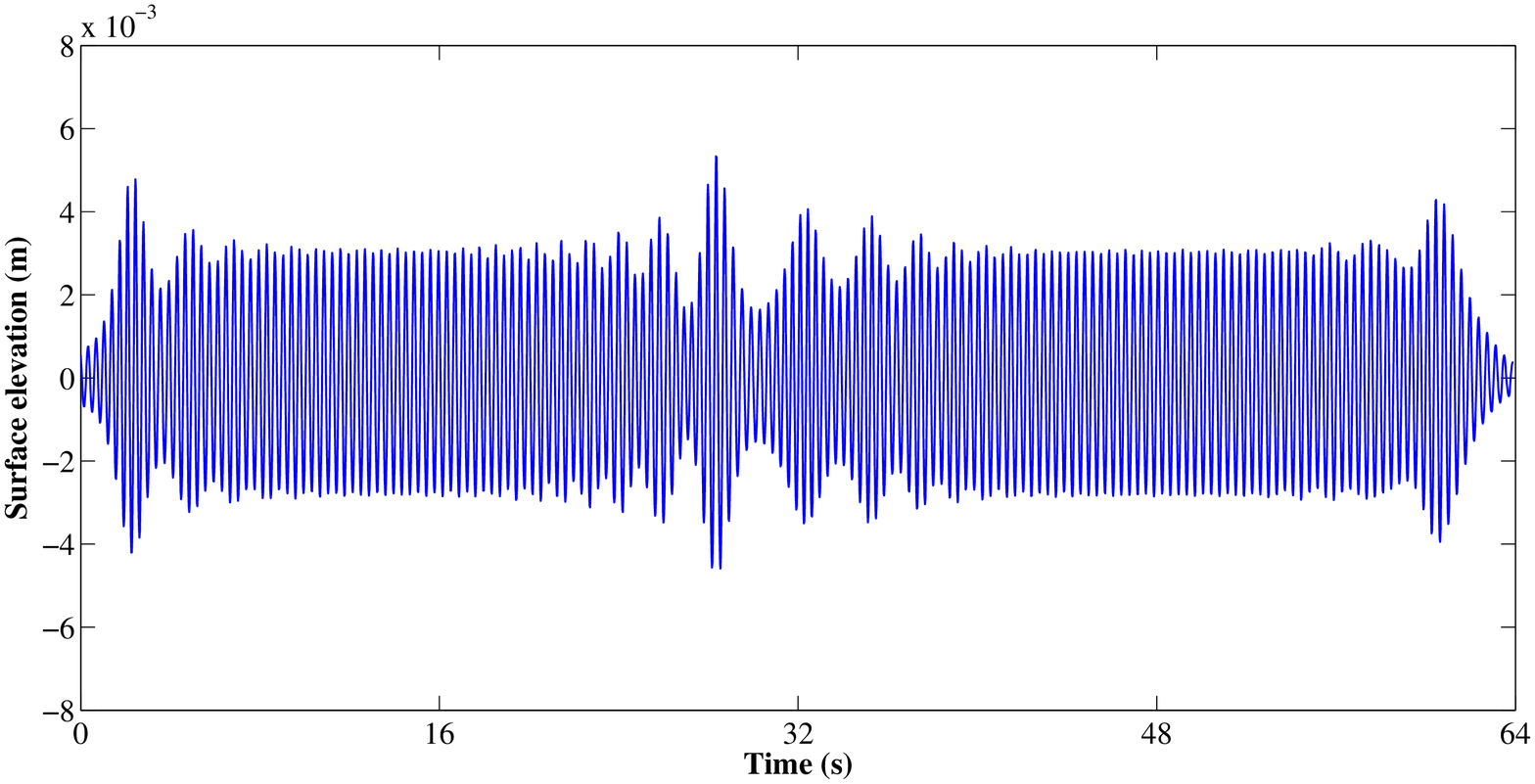}
\includegraphics[width=\linewidth]{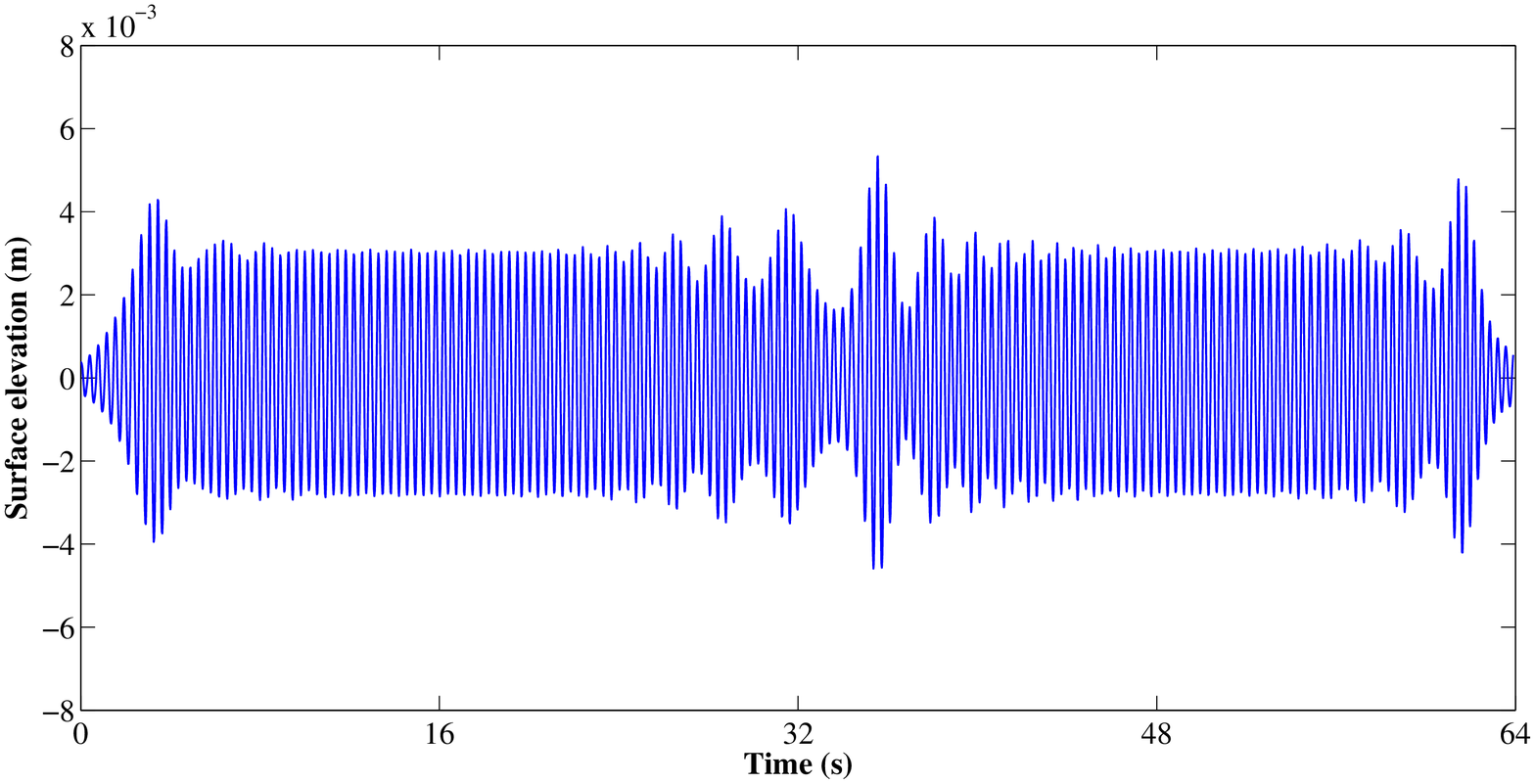}
\caption{Peregrine breather surface elevation with $ka=0.09$ and $a=0.003$ m. (Top) Probe signal measured at $x_M$ after the propagation on a distance $kx_M=270$. (Bottom) Time-reversed signal, providing new wave maker's boundary condition.}
\label{fig:FP_ka_0.09_kx_270}
\end{center}
\end{figure}

Once the first wave field is recorded at the specified probe mirror location $x_M$, we time-reverse this signal and use it as novel wave maker's signal (still using the same linear transfer function) for the follow-up TR refocusing experiment. Figure \ref{fig:FP_ka_0.09_kx_270} presents an example of recorded signal in the case of Peregrine breather as well as the corresponding time-reversed time-series, injected to the wave maker for $ka=0.09$ and $kx_M=270$. Note that the time origin has been shifted to take into account the propagating distance ($t'=t-x_M/C_g$).

\begin{figure}[h!tbp]
\begin{center}
\includegraphics[width=\linewidth]{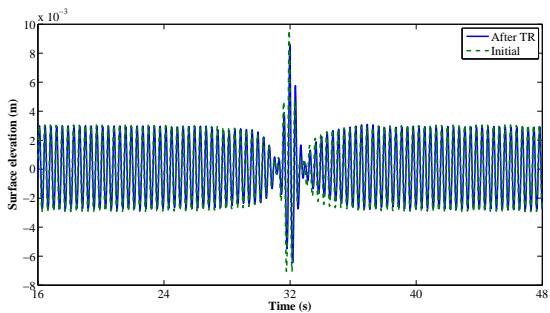}
\caption{Comparison of the Peregrine surface profile as generated by the wave maker (dashed line) with the reconstructed and refocused waves after the TR procedure (solid line). The carrier parameters are $ka=0.09$ and $a=0.003$ m, while $kx_M=270$.}
\label{fig:Peregrine_TR_ka_0.09_kx_270}
\end{center}
\end{figure}

Finally, thanks to the spatial reciprocity the time-reversed and second wave field is generated at the same source and recorded at the same mirror location $x_M$. This allows to compare the focused wave profile to the original signal, as shown in Figure \ref{fig:Peregrine_TR_ka_0.09_kx_270}.
Details about the comparisons and validations with respect to the results presented in \cite{chabchoub2014time} are provided in the Supplemental Material \cite{supplement}. The case of the doubly-localized Peregrine breather and second-order Akhmediev-Peregrine breather are validated. It shows that the proposed numerical procedure, based on the use on the highly nonlinear HOS-NWT provide an accurate description of the complex physics involved in the wave generation and propagation. The next step is now to make use of the proposed procedure to study the application range of the TR in the context of water waves with respect to the steepness (nonlinearity) and the propagation distance (dispersion) for an accurate TR refocusing.

\section{Validity of the time-reversal approach}

Realistic sea state conditions may exhibit configurations, in which nonlinearities may become significant \cite{kharif2009rogue}. Taking into account this fact as well as the chaotic nature of ocean waves, it is consequently interesting to study the possible limitations to the use of TR for those wave conditions. This would allow to determine the limitations in tracing back oceanic extreme events.  

In this study, we reduce our attention to simple stationary as well as highly nonlinear pulsating states, as determined by the NLS. The parameters studied are reduced to the wave steepness $ka$, which characterizes the nonlinearity of the wave field, as well as the maximal propagation distance required for an accurate TR refocusing $kx_M$, that characterizes the effects of dispersion. As simple artificial sea state configurations, we will consider the stationary envelope soliton solution, followed by the two doubly-localized breathers of first and second-order, respectively.

It appears necessary to use an accuracy estimator to evaluate the validity of the TR technique. This is based on the free surface elevation recorded by the wave gauge after application of TR, namely $\eta_{TR}(t)$. In the context of the reproduction of extreme waves in wave basins, it has been shown in \cite{ducrozet2016equivalence} that the most important parameter to ensure the correct kinematics inside the fluid domain, also essential for example to study wave-structure interactions, is the wave amplitude. Consequently, the chosen accuracy parameter is defined by
\begin{equation}
	R_{amp} = \dfrac{ \max \left(\eta_{TR}\right)}{\max \left(\eta_{NLS}\right)},
\end{equation}

\noindent
while $\eta_{NLS}$ denotes the analytic free surface elevation obtained from NLS \eqref{eq:reconstruction_FS}. This expression includes the second-order bound waves which contribution becomes more distinct when strong localized focusing in the wave train emerges. This parameter appears appropriate to have an accurate estimation of the quality of the TR refocusing. As we will see in the following, discrepancies will be characterized by possible change of shape or time-shifts. Nevertheless, these discrepancies have been always associated to a change on amplitude that is relevant as an indicator of global accuracy.

Note that the maximum steepness, chosen in the different test models presented, is below breaking-steepness thresholds, related to each considered solution. Indeed, choosing large steepness values for the carrier, modeled by breathers in particular may engender breaking, making these wave trains are neither suitable for accurate studies within the context of the NLS nor for the HOS-NWT based simulations. We therefore limit this study to the range of stability of experiments and numerical computations.

\subsection{Stationary envelope solitons}

\begin{figure}[h!tbp]
\begin{center}
\includegraphics[width=\linewidth]{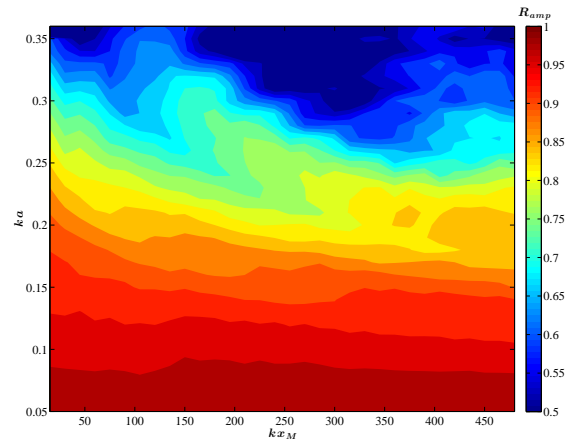}
\caption{Accuracy diagram of the TR for the stationary envelope soliton $R_{amp}$ as function of initial steepness $ka$ and propagating distance $kx_M$.}
\label{fig:sech_TR_accuracy_amplitude}
\end{center}
\end{figure}

We present here the application of the TR refocusing approach to the case of envelope solitons, varying the initial steepness $ka$ and dimensionless propagating distance $kx_M$ accordingly. Figure \ref{fig:sech_TR_accuracy_amplitude} presents the accuracy estimator $R_{amp}$ as function of these two parameters.

First, it is obvious that the accuracy of the TR method is indeed dependent on the two parameters of interest in this study, allowing the quantification of nonlinearities as well as dispersive properties of the considered wave field. We observe that increased nonlinearity of the wave field leads to a less accurate reproduction of the target wave field. These high-order effects start to be very important from $ka \simeq 0.20$ and have a significant influence on the quality of the procedure above this value. Furthermore, the correlation amplitude factor $R_{amp}$ of the considered "sech"-wave pattern starts to drastically reduce, when approaching the wave breaking limit. Note that  experimental results dealing with the one way propagation of such wave field (no TR procedure), reported in \cite{slunyaev2013simulations}, show that at a steepness of $ka=0.35$ strong distortions of the NLS envelope soliton are noticed.

At the same time, also clearly noticeable, larger propagating distances also decrease the accuracy of the TR procedure accuracy. The propagating distances studied here are ranging from $kx_M=15$ up to $kx_M=480$, representing indeed significant long evolution distances. High-order dispersive and nonlinear effects may consequently have indeed an influence on the TR procedure (\emph{i.e.} generation, propagation, measurement of the wave field followed by the refocusing), breaking down the time-reversibility in the proposed configuration.\\

\begin{figure}[h!tbp]
\begin{center}
\includegraphics[width=\linewidth]{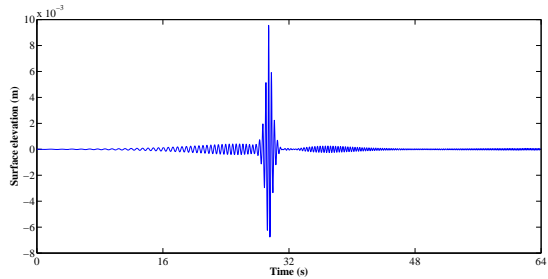}
\caption{Free surface elevation of the envelope soliton for an amplitude $a=0.01$ m with steepness of $ka=0.30$, recorded at the probe after a dimensionless propagation distance of $kx_M=330$.}
\label{fig:sech_FP_kx_330_ka_0.3}
\end{center}
\end{figure}

In order to refine the analysis of the origin of the discrepancies observed between initial and reconstructed free surface elevation, we present the simulation results, obtained for the steepness $ka=0.30$, an amplitude of $a=0.01$ m and propagating distance $kx_M=330$. The corresponding amplitude ratio in this case is $R_{amp}=0.54$, indicating low accuracy of the TR procedure with this choice of parameter. Figure \ref{fig:sech_FP_kx_330_ka_0.3} shows the corresponding results after the first propagation in the NWT. 

\begin{figure}[h!tbp]
\begin{center}
\includegraphics[width=\linewidth]{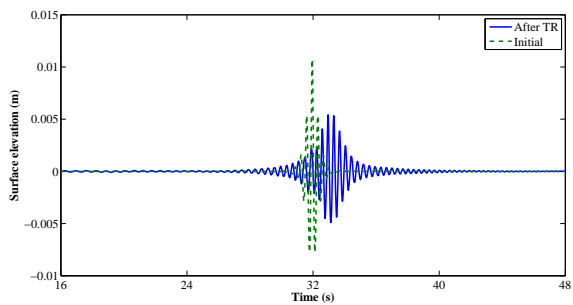}
\caption{Results of Time Reversal for an envelope soliton for an amplitude $a=0.01$ m, steepness $ka=0.30$ and $kx_M=330$.}
\label{fig:sech_TR_kx_330_ka_0.3}
\end{center}
\end{figure}

We can observe that the initial envelope soliton is starting to disintegrate during its propagation and its dynamics is not in agreement with NLS predictions. Again, this is in agreement with \cite{slunyaev2013simulations}. The initial "sech"-shaped solution starts to fission into smaller amplitude solitons \cite{clamond2006long}. Figure \ref{fig:sech_TR_kx_330_ka_0.3} depicts the comparison of the initial envelope soliton and the free surface elevation, reconstructed through TR. 

We notice that neither the initial amplitude nor the initial location could be reconstructed. The initial deterioration of the envelope after the first propagation is exacerbated by the TR after the second evolution. This means that from the first wave measurement (Fig. \ref{fig:sech_FP_kx_330_ka_0.3}), the TR procedure which deduces a new wave maker's signal and a second propagation is inducing some discrepancies that breaks down the time-reversibility. In fact, significant higher-order effects in terms of nonlinearities and dispersion are thus associated to the reduced accuracy. Those features will be studied in more details with the analysis of doubly-localized breathers.

\subsection{Doubly-localized breathers}

The study of limitations of the TR approach for breathers is obviously more challenging, compared to the previous stationary case. The first- and second-order doubly-localized breathers have the property to have an infinite modulation period, exhibiting a strong variation of wave envelope and therefore of spectra as well. Detecting the limitations of applicability of the TR procedure may be transferred to assigning limitations with respect to the reconstruction of extreme event in the ocean within the framework of integrable evolution equations.

As already mentioned earlier, the accuracy of the NLS and applicability of the TR technique is closely linked to the initial steepness $ka$ of the carrier. A similar dependence to the propagating distance $kx_M$ can be also expected.

\subsubsection{Peregrine}

\begin{figure}[h!tbp]
\begin{center}
\includegraphics[width=\linewidth]{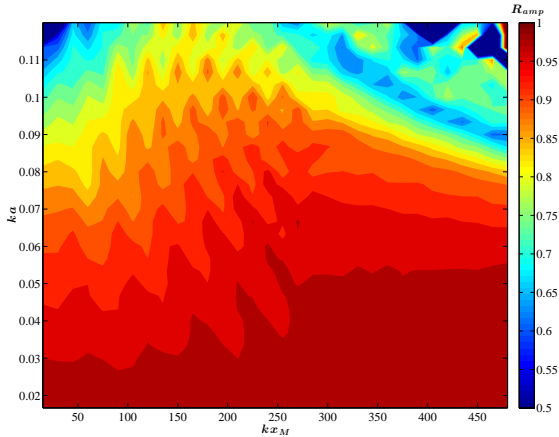}
\caption{Accuracy diagram of the TR for the doubly-localized Peregrine breather $R_{amp}$ as function of initial steepness $ka$ and propagating distance $kx_M$.}
\label{fig:Peregrine_TR_accuracy_amplitude}
\end{center}
\end{figure}

We propose here some complements to Figure \ref{fig:Peregrine_TR_ka_0.09_kx_270}, which depicted a result of the TR method for a given steepness $ka=0.09$ and propagating distance $kx_M=270$ in the case of a Peregrine breather. Figure \ref{fig:Peregrine_TR_accuracy_amplitude} presents the accuracy of the refocusing after the TR in terms of amplitude of the extreme wave ($R_{amp}$). The parameters of interest are varied in a range $kx_M \in \left[ 15 ; 480 \right]$ and $ka \in \left[ 0.02 ; 0.12 \right]$. Reminding that the theoretical amplification of plane wave for a Peregrine breather is of three, this choice of steepnesses covers the whole range of existence of such localized structures, below the breaking threshold, experimentally determined of being at a steepness value of about 0.12 \cite{chabchoub2012experimental}.

It is also confirmed in this case that the accuracy of the approach is improved, when the steepness is lower. In fact, we can ensure an accurate validity of TR refocusing for the whole range of dimensionless distance $kx_M$, for steepness values below $ka<0.075$. Here, we also emphasize a slight decrease of accuracy in the latter range for small propagation distances $kx_M < 100-150$.

Note that Figure \ref{fig:Peregrine_TR_accuracy_amplitude} allows to extract the accuracy of laboratory breather TR experiments, as described in \cite{chabchoub2014time}. Indeed, for the chosen propagating distance $kx_M=270$ the TR for the Peregrine breather with good exactitude applies up to a steepness $ka \simeq 0.09$ (consistent with the successful experiments in \cite{chabchoub2014time} conducted for $ka=0.09$).
We recall that the local steepness of the largest wave in the group is indeed very large and can be roughly estimated of being three time the carrier steepness, that is of approximately $(3~ka) \simeq 0.27$. Taking into account the theoretical limitations of the practical TR procedure, as well as the laboratory or numerical noise, always present in the experiment, the validity of TR approach for this range of parameters is remarkable. 

As stated previously, beyond $ka>0.075$, a good accuracy of TR refocusing can still noticed for some values of steepness and dimensionless propagation distance. However, this validity range is not as obvious. This will be studied in more details in Sec. \ref{sec:physical_interpretation}.

\subsubsection{Akhmediev-Peregrine}

In the case of second-order doubly-localized breather, validations of the TR refocusing are provided in the Supplemental Material \cite{supplement} for $ka=0.03$ and $kx_M=270$.
Figure \ref{fig:Peregrine_TR_accuracy_amplitude} extends this study and presents a more general accuracy diagram with respect to the parameter $R_{amp}$, as function of initial steepness $ka$ and propagating distance $kx_M$, as in the previous case. The range of parameters is $kx_M \in \left[ 15 ; 480 \right]$ and $ka \in \left[ 0.01 ; 0.072 \right]$. We have to recall that for this specific solution, the theoretical amplification of plane wave is of five. Again, the range of steepness values has been chosen to prevent breaking of the highest amplified waves.

\begin{figure}[h!tbp]
\begin{center}
\includegraphics[width=\linewidth]{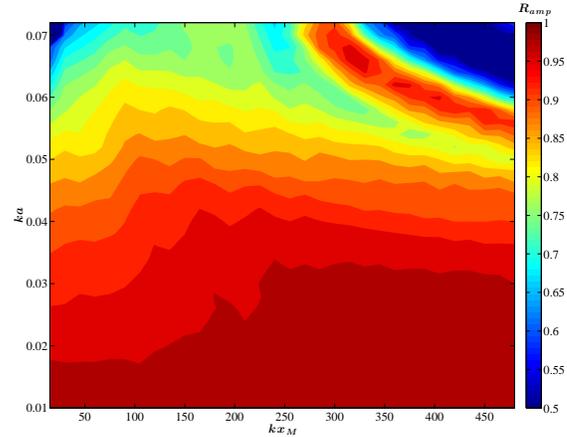}
\caption{Accuracy diagram of the TR for the doubly-localized Akhmediev-Peregrine breather of order two $R_{amp}$ as function of initial steepness $ka$ and propagating distance $kx_M$.}
\label{fig:Peregrine_2_TR_accuracy_amplitude}
\end{center}
\end{figure}

The dependence of $R_{amp}$ with respect to the two parameters is very similar to the case of the Peregrine breather. That is to say, the accuracy of the TR approach is reduced when the steepness is increased. At the same time, the accuracy is also reduced at large propagating distances, and slightly altered for $kx_M < 100-150$. Figure \ref{fig:Peregrine_TR_accuracy_amplitude} may be used to state that for the whole range of propagating distances studied, the TR method allows correct refocusing up to $ka \simeq 0.045$.
For the the same example than in previous section (\textit{i.e.} $kx_M = 270$), the TR method allows correct refocusing up to $ka \simeq 0.055$, that is at a local steepness in the modulation $(5~ka) \simeq 0.28$.\\

As a general conclusion of the previous test-cases, it has been shown that the TR method is efficient on a large range of steepnesses and propagating distances. Different kinds of wave envelope patterns have demonstrated a similar behavior. However, the practical set-up of time-reversibility of water waves fails for large steepness or very large propagating distances. TR procedure will allow an accurate refocusing when the local steepness of the localized structures do not exceed a general value of about $ka \simeq 0.25$. Nevertheless, we are optimistic in possible applications of TR for ocean waves, considering the fact that ocean waves have the property to present reduced steepness values most of the time. 

More physical insights with respect to the TR approach will be provided in the next section. 

\section{Physical interpretation of the HOS-NWT simulations} 
\label{sec:physical_interpretation}

This section is devoted to the analysis of the limitations of the TR method, as reported in the previous one. For the sake of brevity, we focus on one particular case, namely the Peregrine breather for the steepness parameter $ka=0.09$, propagating forth and back for $kx_M=270$. We provide a rigorous analysis of the possible influence of the propagating distance, and then of the steepness. Note that essentially similar results are obtained in the case of Akhmediev-Peregrine breather.

\subsection{Influence of propagating distance $kx_M$}
\label{subsec:physical_interpretation_kx}

As already explained previously, one source of discrepancies is associated to the practical set up of TR procedure in the context of water waves. The wave maker imposes the only free surface elevation and is deduced from linear theory. This will influence the way the breather will propagate and consequently it is awaited that the effects will become increasingly significant with the increase of the propagating distance. This is the probable origin of the behavior observed in Figs. \ref{fig:Peregrine_TR_accuracy_amplitude} \& \ref{fig:Peregrine_2_TR_accuracy_amplitude} at large distances of evolution.

\begin{figure}[h!tbp]
\begin{center}
\includegraphics[width=\linewidth]{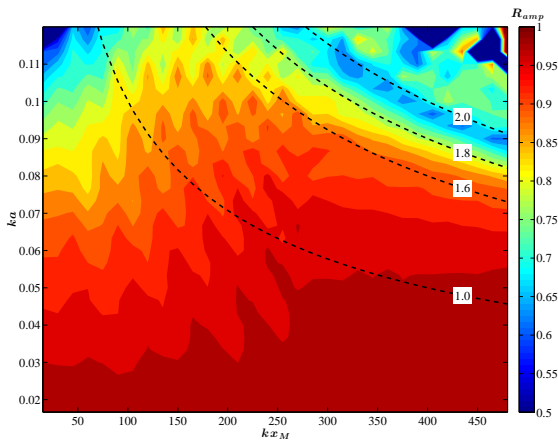}
\caption{Accuracy of the TR method for the Peregrine breather. Dashed lines are iso-lines in terms of the modulation instability process.}
\label{fig:Peregrine_TR_accuracy_amplitude_with_BF_instability}
\end{center}
\end{figure}

Note that the time scale of Benjamin-Feir (BF) instability is determined by $t_{BF} = \mathcal{O}\left({T_0}/{(ka)^2}\right)$. The corresponding length scale of this modulation instability is $kx_{BF} = k C_g t_{BF} = \mathcal{O}\left( {1}/{(ka)^2}\right)$, with $C_g$ the group velocity. It is known that this BF instability is at the origin of the possible existence of localized structures such as breathers \cite{osborne2010nonlinear}. In the present case ${1}/{(ka)^2} \simeq 125$ and consequently the propagating distances studied are in the range of modulation instability processes' scale. This also indicates that there is indeed a strong link between the propagating distance $kx_M$ and the steepness of the considered wave field $ka$. In complement, Figure \ref{fig:Peregrine_TR_accuracy_amplitude_with_BF_instability} presents for the previous accuracy diagram iso-lines corresponding to this BF instability in the range $kx_{BF} \in \left[ 1 ; 2 \right]$. 

This instability appears consequently as a major point in the understanding of the limitations inherent to the TR method. In the case of a Peregrine breather, when propagating distance is larger than about $1.8$ times the length scale of BF instability, the applicability of the TR method starts to fade. This can be related to the scale of evolution of the breather solution to NLS, and of development of the possible perturbations to this solution.

\begin{figure}[h!tbp]
\begin{center}
\includegraphics[width=\linewidth]{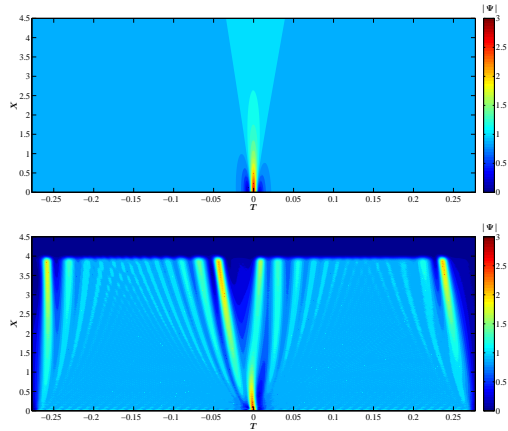}
\caption{Propagation of the Peregrine breather ($ka=0.09$) in the wave basin, space-time representation in scaled NLS variables. Comparison of analytic solution (Top) and HOS-NWT numerical simulation (Bottom).}
\label{fig:Peregrine_FP_2dpt_ka_0.09}
\end{center}
\end{figure}

In addition to the free surface time signals, numerical simulations enable a general overview of the wave field evolution. Figure \ref{fig:Peregrine_FP_2dpt_ka_0.09} presents a full space-time depiction of the propagation inside the wave basin. The variables are changed to NLS scaled variables 
$X = \sqrt{2}k^2a x$ and $T=-\dfrac{k^2a^2\omega}{4} \left(t - x/C_g\right)$
 as a matter of comparison to analytic solution. We remind that an absorbing zone is present to prevent from wave reflections on the end wall in the HOS-NWT simulations. This appears clearly on Figure \ref{fig:Peregrine_FP_2dpt_ka_0.09} at $X>4$.

The demodulation of the initial localized structure with respect to the propagating distance is obvious during its evolution. However, it is also worth to note that the solution does not tend to the plane wave solution after a long time/distance propagation for this choice of carrier parameters, as expected from NLS theory. Note that it has been verified that this is not related to the finite extent in time of the generated wave train.

At the same time, an asymmetry of the temporal wave probe signal can also be observed after the first propagation. Indeed, the NLS equation predicts a symmetric demodulation with respect to time, at fixed spatial location, which appears to be erroneous when nonlinearities are important. This fact is well-known and has already been pointed out in different studies. It has been shown that higher-order NLS equation, also referred to as the Dysthe equation, enables a non-symmetric prediction of the envelope evolution \cite{trulsen1996modified,lo1985numerical,slunyaev2013super,chabchoub2013hydrodynamic,shemer2013peregrine}, when the steepness of the carrier and the amplitude amplification become substantial.

\begin{figure}[h!tbp]
\begin{center}
\includegraphics[width=\linewidth]{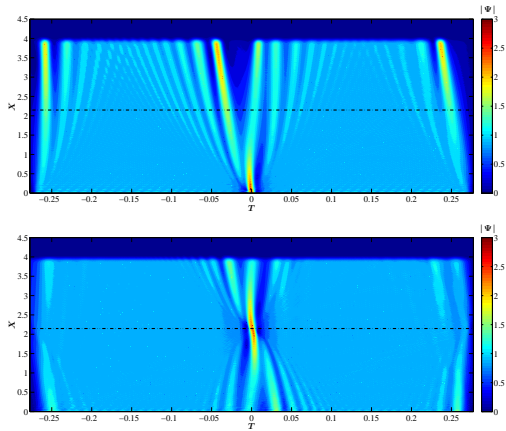}
\caption{Space and time evolution of Peregrine breather for $ka=0.09$ and $kx_M=270$ (dash-dotted line) in the NWT. (Top) Demodulation (Bottom) TR refocusing.}
\label{fig:Peregrine_FP_and_TR_2dpt_kx_270_ka_0.09}
\end{center}
\end{figure}

Nevertheless, as seen previously in Figure \ref{fig:Peregrine_TR_ka_0.09_kx_270}, the noticed and expected asymmetry does not influence the TR invariance. Figure \ref{fig:Peregrine_FP_and_TR_2dpt_kx_270_ka_0.09} depicts the space and time evolution of the Peregrine wave field during the first propagation as well as for the TR wave refocusing. The case of the Peregrine breather for $ka=0.09$ and the propagating distance of $kx_M=270$ is adopted here. This corresponds to a scaled distance $X_M \simeq 2.1$ depicted on the figure.

The time-reversibility is clearly demonstrated during the whole HOS-NWT computation, throughout the whole spatial domain. Again, this confirms the accurate refocusing presented in Figure \ref{fig:Peregrine_TR_ka_0.09_kx_270} and in the experiments in \cite{chabchoub2014time}. This point is of major importance since even if the wave field evolution clearly departs from NLS prediction, the full non-linear system appears as time-reversible. The TR procedure enables the use of such property of the wave field, even in the case of high nonlinearity.\\

\begin{figure}[h!tbp]
\begin{center}
\includegraphics[width=\linewidth]{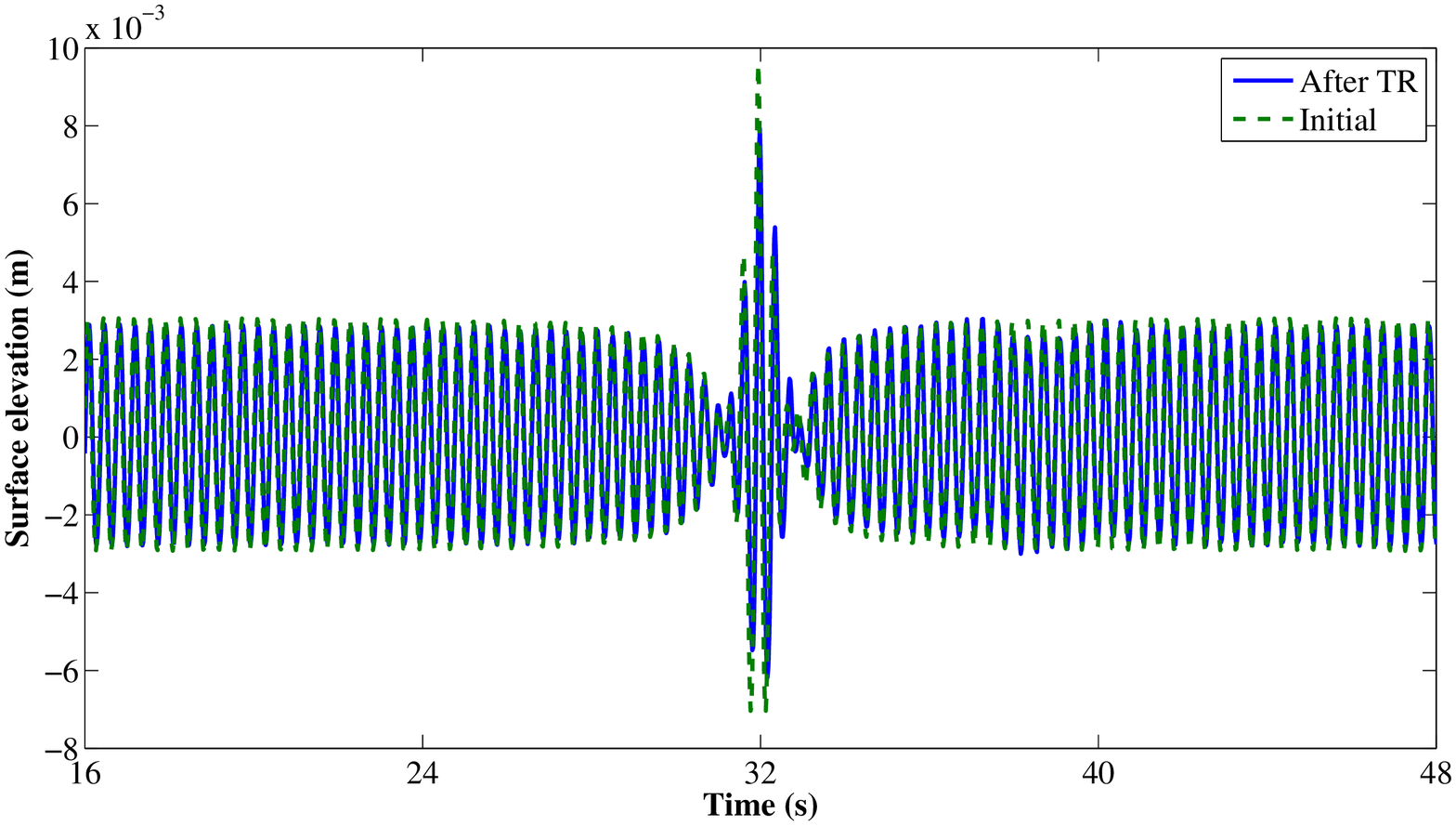}
\includegraphics[width=\linewidth]{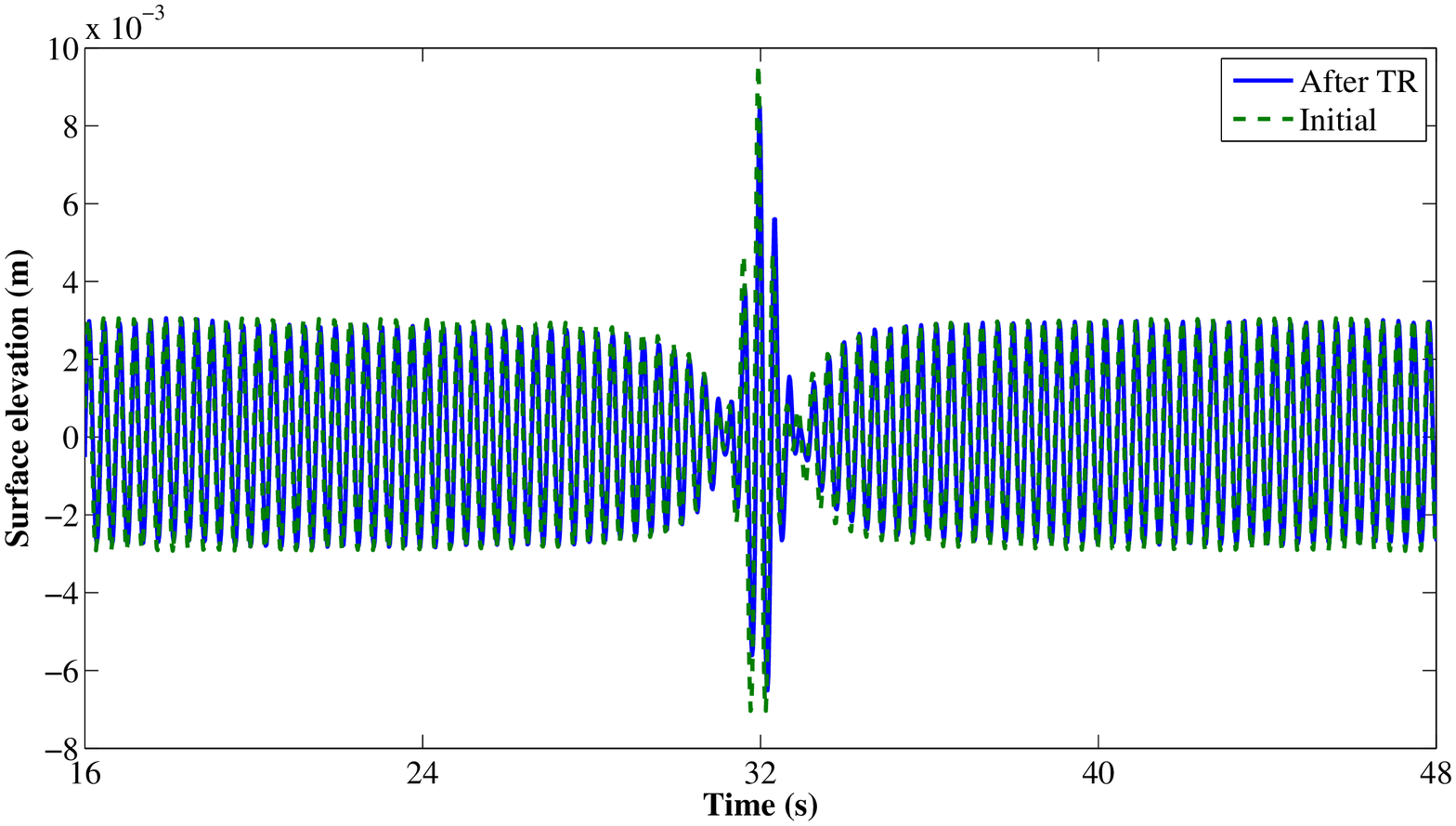}
\includegraphics[width=\linewidth]{Peregrine_Time_reversal_kx_270_ka_0.09_comp_2nd.eps}
\includegraphics[width=\linewidth]{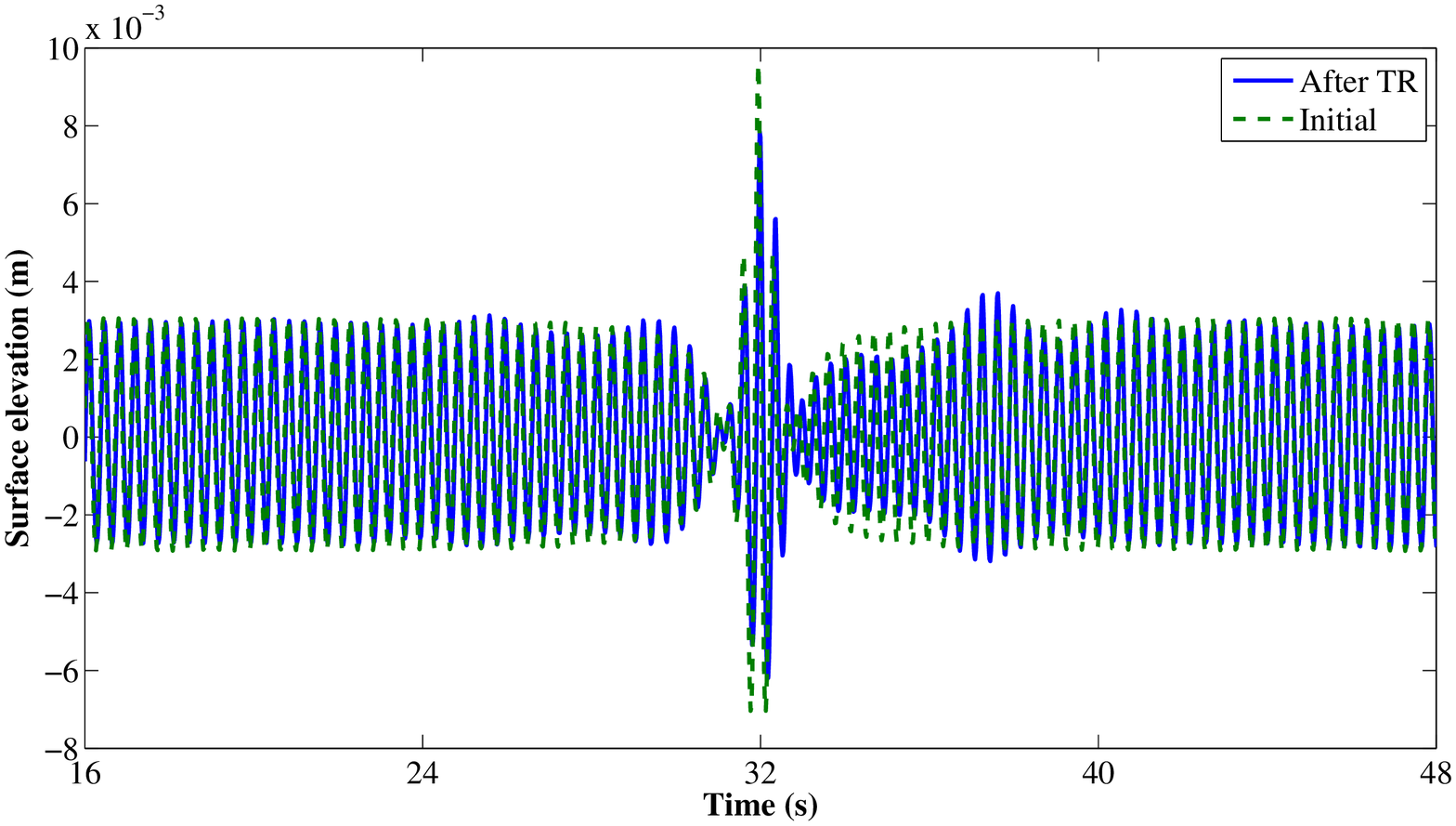}
\includegraphics[width=\linewidth]{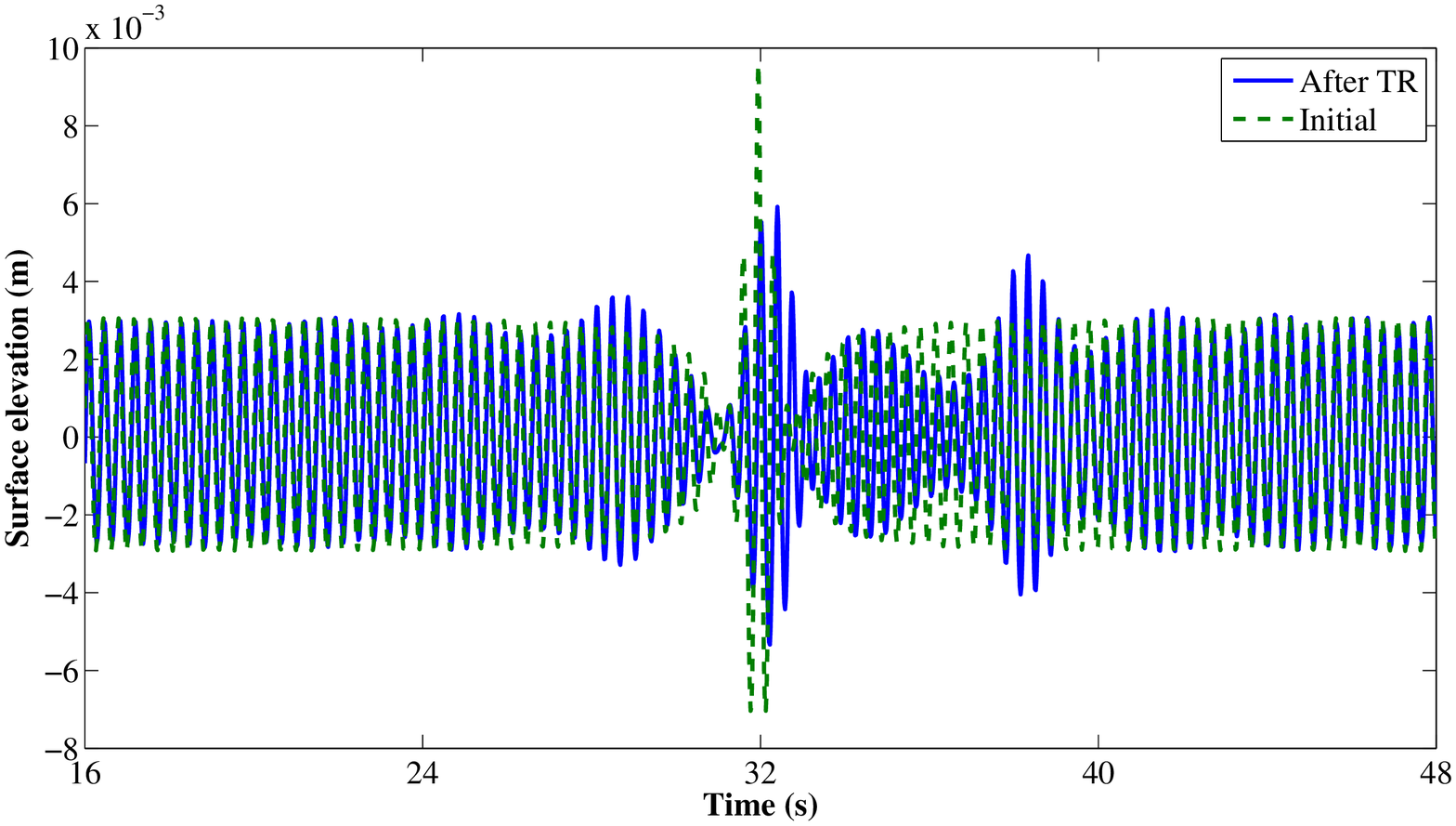}
\caption{Comparison of the Peregrine surface profile at maximal compression, as initially generated with respect to the NLS theory (dashed line) and after TR refocusing (solid line), for $ka=0.09$ and $kx_M\in\{60;165;270;375;480\}$ (from Top to Bottom).}
\label{fig:Peregrine_TR_ka_0.09}
\end{center}
\end{figure}

The influence of the propagating distance is now studied on the previous test-cases by gradually increasing the values of $kx_M$. Figure \ref{fig:Peregrine_TR_ka_0.09} shows the final refocusing results of the complete TR procedure for $kx_M\in\{60;165;270;375;480\}$ at a fixed steepness of $ka=0.09$.

We can observe in this latter figure that the influence of the propagating distance on the reconstruction is quite complex. For relative small propagating distances ($kx_M<100$), the global shape of the localized structure is correct but is associated to a reduced amplitude at the location of the extreme wave. Then, for a significant range $kx_M \in \left[ 100 ; 350 \right]$, the TR procedure is very accurate in the refocusing of the localized structure of interest. The influence of the propagating distance is then to induce an increasing but still very small phase-shift between the refocused solution and the original one. Finally, for very large evolution distances, starting from $kx_M \simeq 375$, the slight phase-shift after TR is associated to an ineffective refocusing. This leads to a complex wave field distortion and wave profile that significantly deviate from the initial doubly-localized structure. As seen in Figure \ref{fig:Peregrine_TR_ka_0.09}, the effect is more pronounced when the distance is enlarged.\\


Also, the origin of this distortion is not straight-forward to explain. It is reminded here that two processes of wave generation and propagation have to be successively conducted in order to obtain the final TR result.

The first generation associated to the propagation creates a nonlinear wave field (that can be different from the NLS solution due to high-order effects and/or discrepancies on the wave maker's motion). Then, during the second stage of TR refocusing, some information may be lost when deducing the new wave maker's motion from the measured free surface elevation $\eta(x_M,t)$ (reminding that we use linear theory for this purpose). Consequently, the final phase-shift or change of observed envelope shape is a kind of footprint with respect to the existence of non-reversible effects in the TR procedure. These are related to nonlinearities possibly together with dispersive effects.

The first influence of the propagation distance is the increase of the phase-shift after the use of the TR method. This reduces the accuracy of the refocusing at large distances. Then, those effects are obviously more enhanced at even larger propagation distances and lead to the failure of the TR procedure. As stated previously, we can evaluate that the length-scale of this limitation of applicability is associated to the modulation instability space-scale.\\

Finally, we would like to address the unexpected validity of the TR approach in all temporal signals at smaller propagating distances ($kx_M < 150$), as can be observed in Figure \ref{fig:Peregrine_TR_accuracy_amplitude}. It is known that during the generation of waves by a wave maker, the so-called evanescent waves are spontaneously created when the wave field is generated. These waves are only relevant close to the wave generator since they quickly decay in space. In the configuration tested, they are therefore negligible considering the whole range of propagation distances for $kx_M \geq 15$. In addition, parasitic second-order free waves are also generated with the target wave field  \cite{schaffer1996second}. As a first approximation, if we consider the waves to be regular, this parasitic wave field are regular free waves with a pulsation $\omega_2=2\omega$. Those waves will propagate at a velocity which is half the velocity of the corresponding waves we are interested to generate and reach the probe position of interest at $t_{2}=2 {x_M}/{C_{g}}$. We recall that figures presenting probe signals are already time-shifted with $t_{\operatorname{shift}} = {x_M}/{C_{g}}$. In this time reference frame, the parasitic second-order free waves, generated by the wave maker, will reach the probe at a time $t'_{2} = {x_M}/{C_{g}}$. For the case $kx_M=60$ presented in Figure \ref{fig:Peregrine_TR_ka_0.09}, this corresponds to $t'_{2} \simeq 7s$. It can be seen on the free surface wave profile that after the breather with a delay $t'_{2}$, unexpected discrepancies are present that are associated to the parasitic second order waves.


Thus, these waves are at the origin of the reduced accuracy at relative small propagating distances $kx_M < \left[100 ; 150\right]$ as reported from Figure \ref{fig:Peregrine_TR_accuracy_amplitude}. Indeed, in this range of distance, the parasitic free waves are recorded after the first propagation for the corresponding propagating time is $t'_{2} < \left[ 12 ; 17 \right]$ s. Consequently, during the refocusing of the waves, these are generated again and interfere with the localized structure, reducing the corresponding accuracy as consequence. It has to be noted that it is possible to control the generation of the parasitic waves, produced by the wave maker as explained in \cite{schaffer1996second}. The wave maker's motion can then be modified in order to prevent the generation of those waves. This approach may be used to possibly increase the accuracy of the TR method.

\subsection{Influence of initial steepness $ka$}
\label{subsec:physical_interpretation_ka}


This part focuses on the influence of the nonlinearity on the TR procedure. As observed previously, the Peregrine solution imposed at the wave maker's location does not tend to the plane wave solution after a long time/distance propagation, as expected from NLS, for significant carrier steepness values. 

\begin{figure}[h!tbp]
\begin{center}
\includegraphics[width=\linewidth]{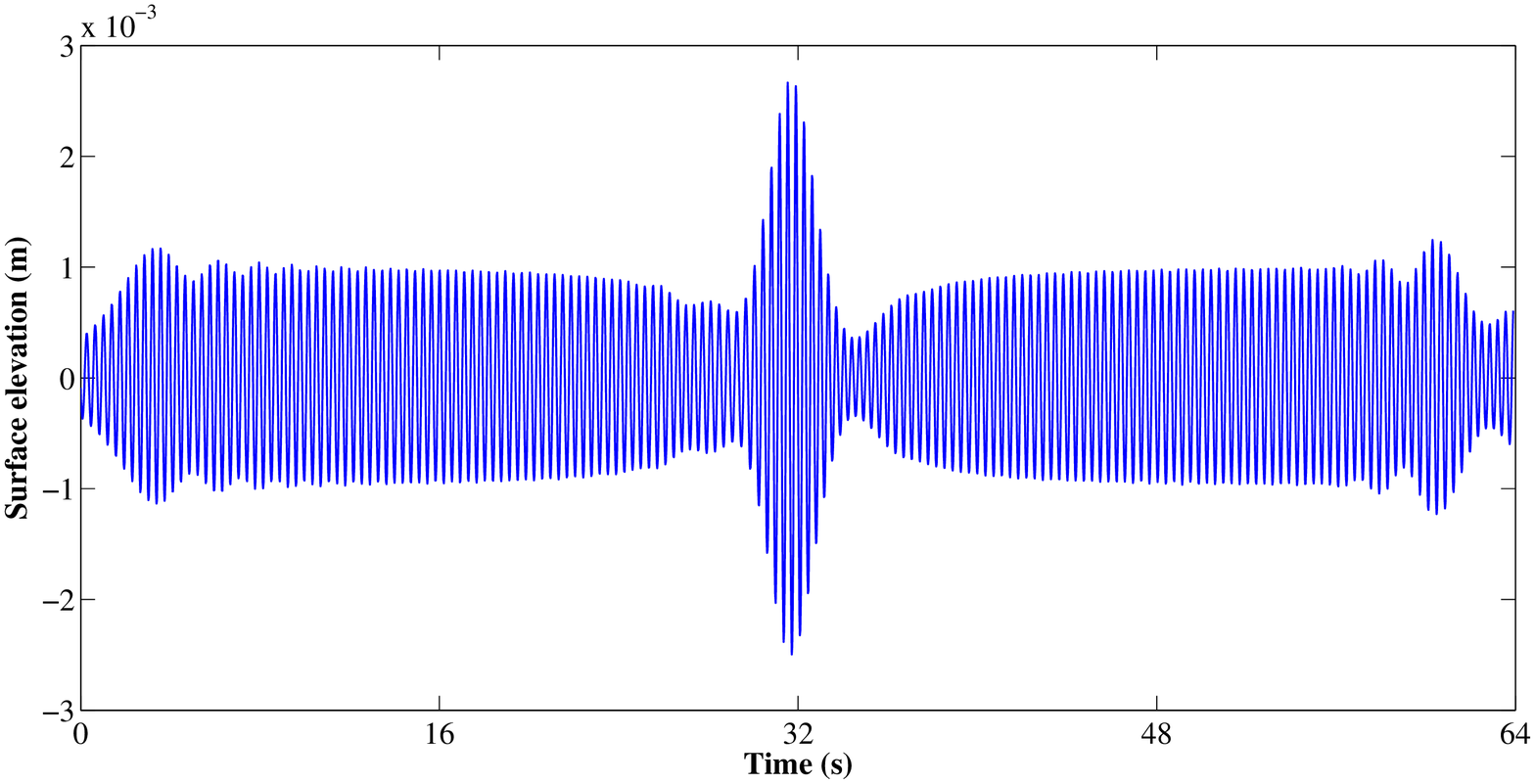}
\includegraphics[width=\linewidth]{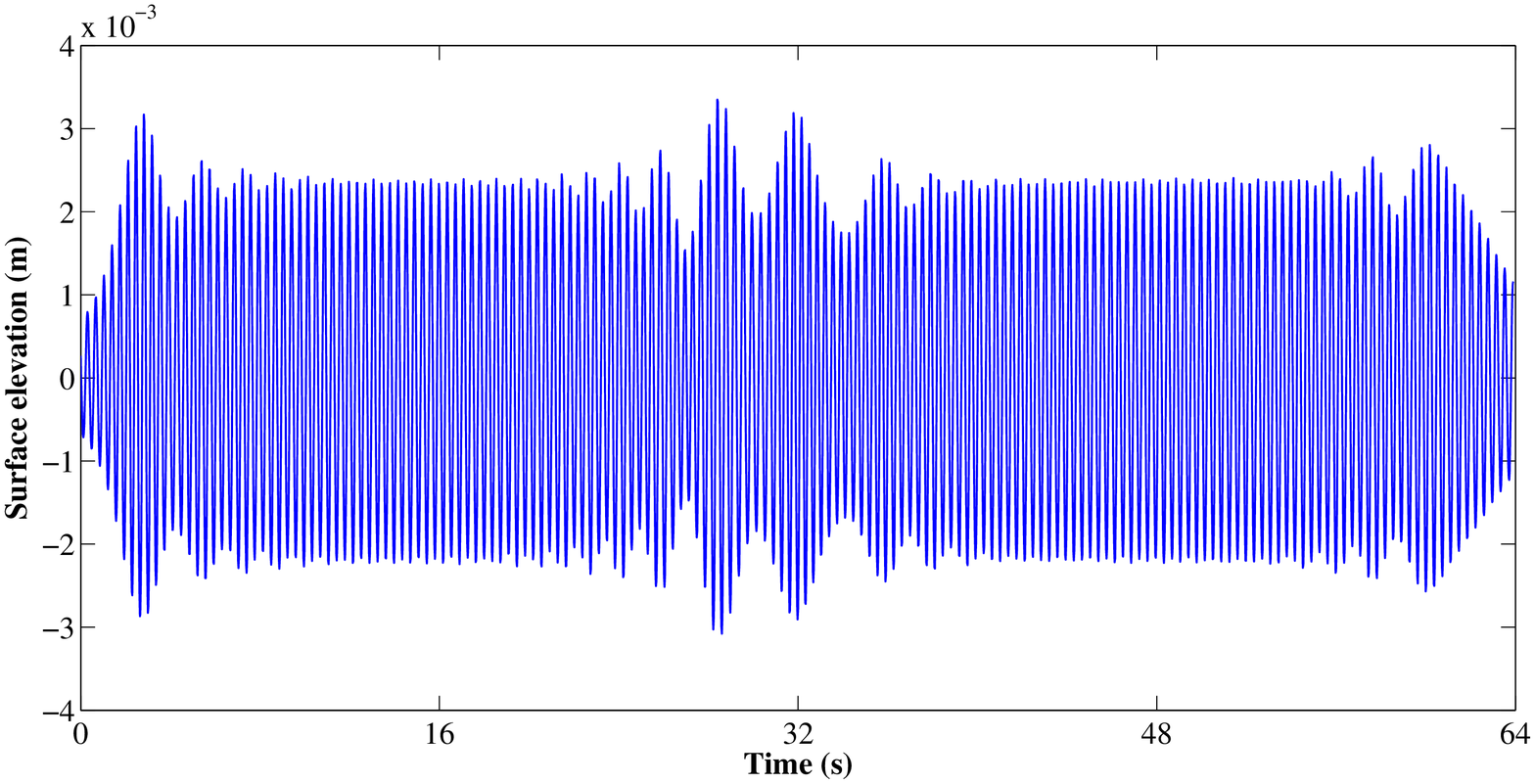}
\includegraphics[width=\linewidth]{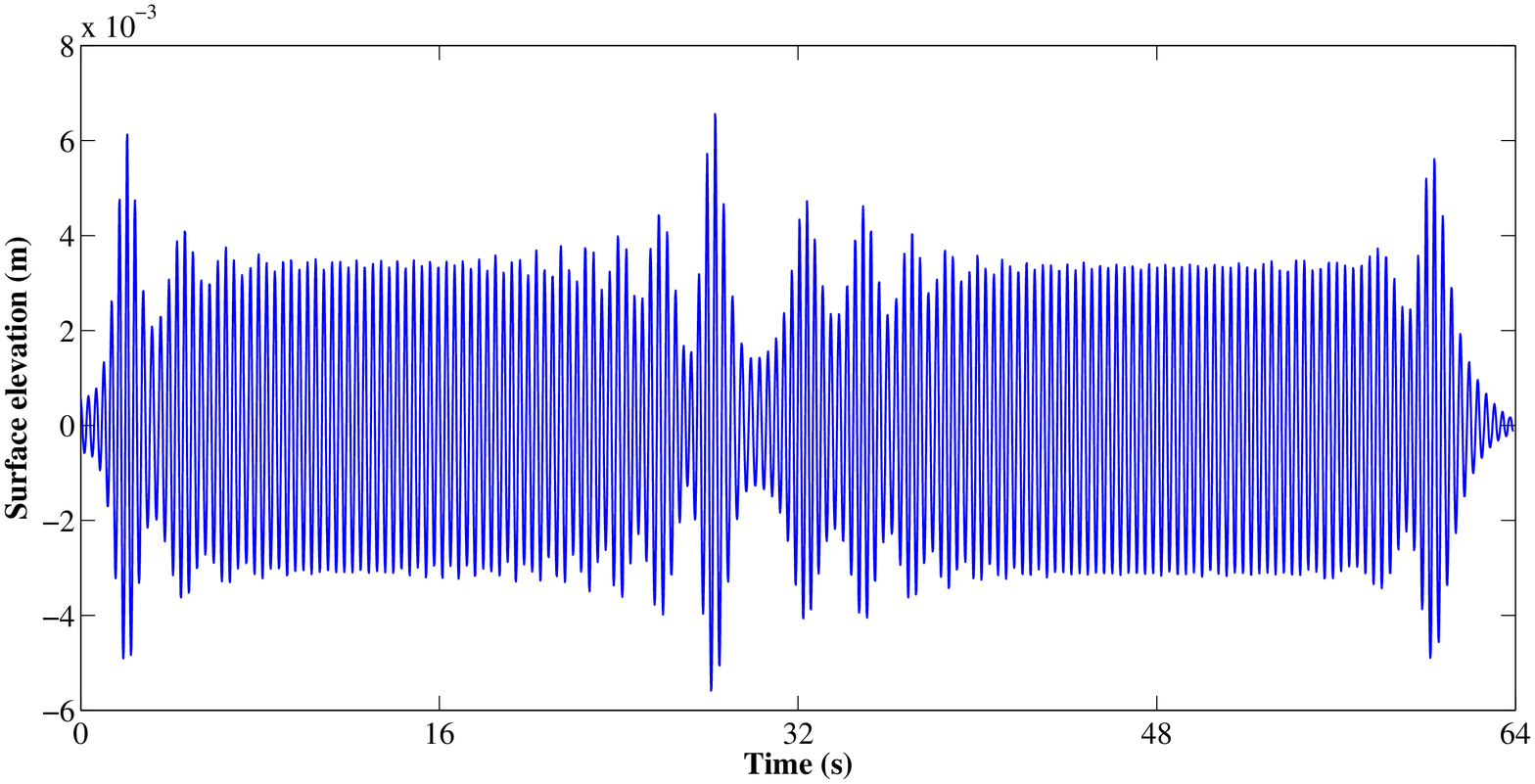}
\includegraphics[width=\linewidth]{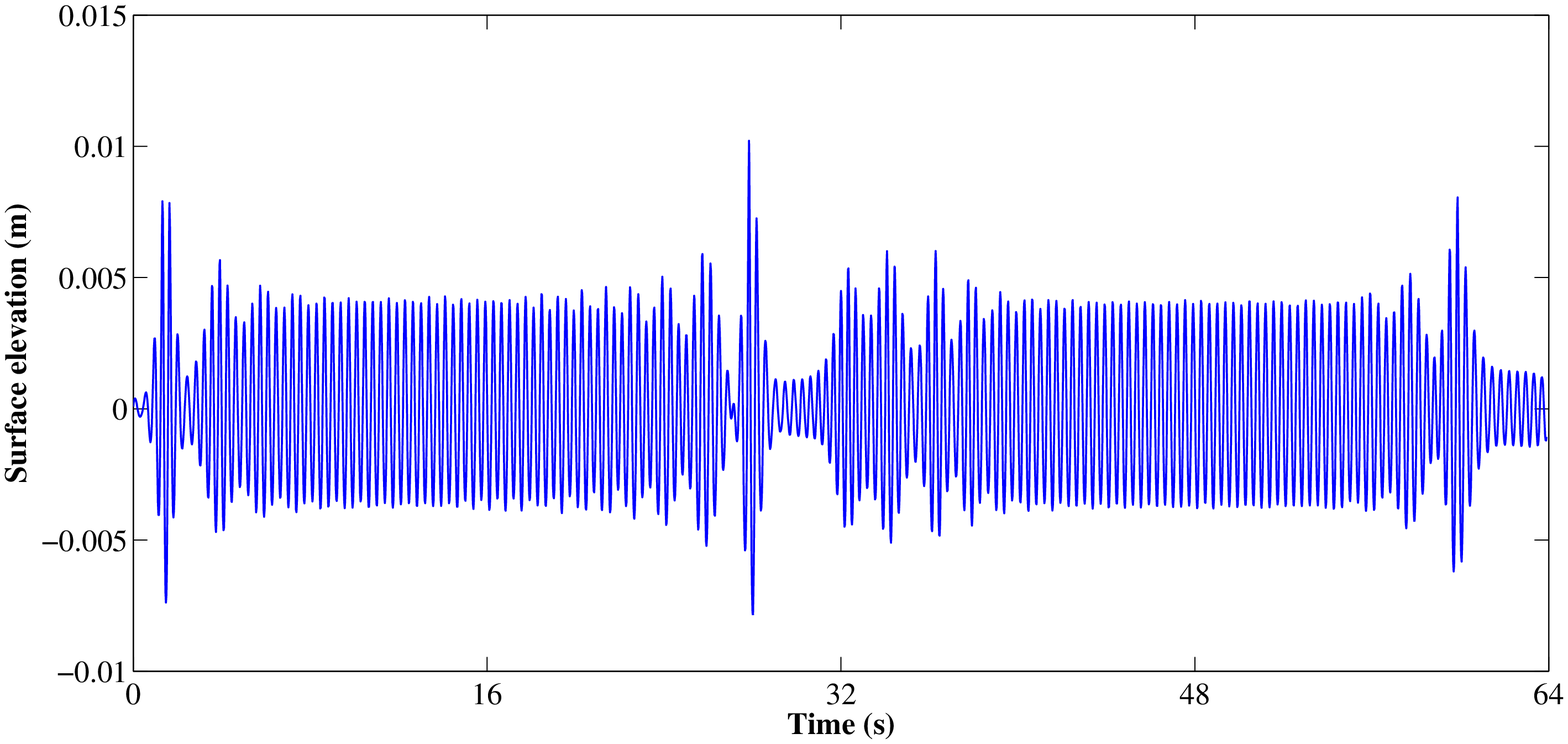}
\caption{Probe signal measured after the propagation of Peregrine breather for $kx_M=270$ for different steepness $ka\in\{0.03;0.07;0.10;0.12\}$ (from Top to Bottom).}
\label{fig:Peregrine_FP_kx_270}
\end{center}
\end{figure}

The previous section presented the influence of the propagation distance $kx_M$ on a TR refocusing accuracy at a fixed steepness $ka$. We will now fix the mirror location and vary the steepness values accordingly. Indeed, we recall that the modulation instability, which drives the evolution of the breather solution, acts on a spatial scale being $kx_{BF} = \mathcal{O} \left( {1}/{(ka)^2}\right)$. Figure \ref{fig:Peregrine_FP_kx_270} presents the probe signals after first propagation of a Peregrine breather at a fixed distance $kx_M=270$ for different initial steepnesses in the range $ka \in \{0.03;0.07;0.10;0.12\}$. 

\begin{figure}[h!tbp]
\begin{center}
\includegraphics[width=\linewidth]{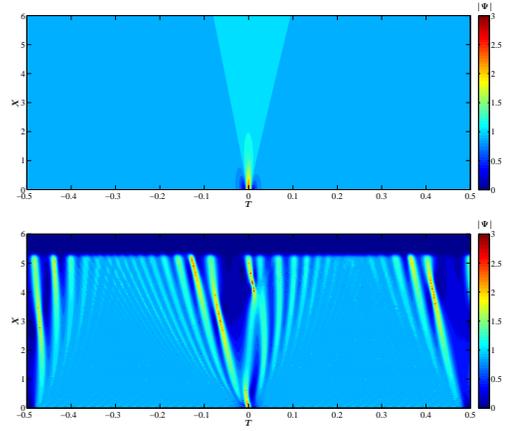}
\caption{Space and time evolution of the Peregrine breather for $ka=0.12$ in the NWT. Comparison of analytic solution (Top) and HOS-NWT numerical simulation (Bottom).}
\label{fig:Peregrine_FP_2dpt_ka_0.12}
\end{center}
\end{figure}

As expected and shown, the relative propagating distance is different with respect to the breather demodulation, evolving on the time scale ${1}/{(ka)^2}$. Consequently, the amplitude decrease of the initial localized structure is highly dependent on this initial steepness \cite{chabchoub2012experimental}. At low steepness, the fixed propagation distance $kx_M$ appears as very small with respect to the modulation distance: the demodulation is less significant and it is expected that the refocusing is more accurate. On the contrary, when the steepness is larger, this chosen distance appears becoming large with respect to modulation instability distance. As a consequence, and due to significance of the nonlinear processes, the solution can appear as highly complex since those high-order effects prevent from recovering the plane wave solution. For details, we refer to Figure \ref{fig:Peregrine_FP_2dpt_ka_0.12} that presents the full space-time plane view of the propagation within the NWT for the steepest case $ka=0.12$, while the variables are changed to NLS scaled variables as a matter of comparison to the scaled analytic solution.

\begin{figure}[h!tbp]
\begin{center}
\includegraphics[width=\linewidth]{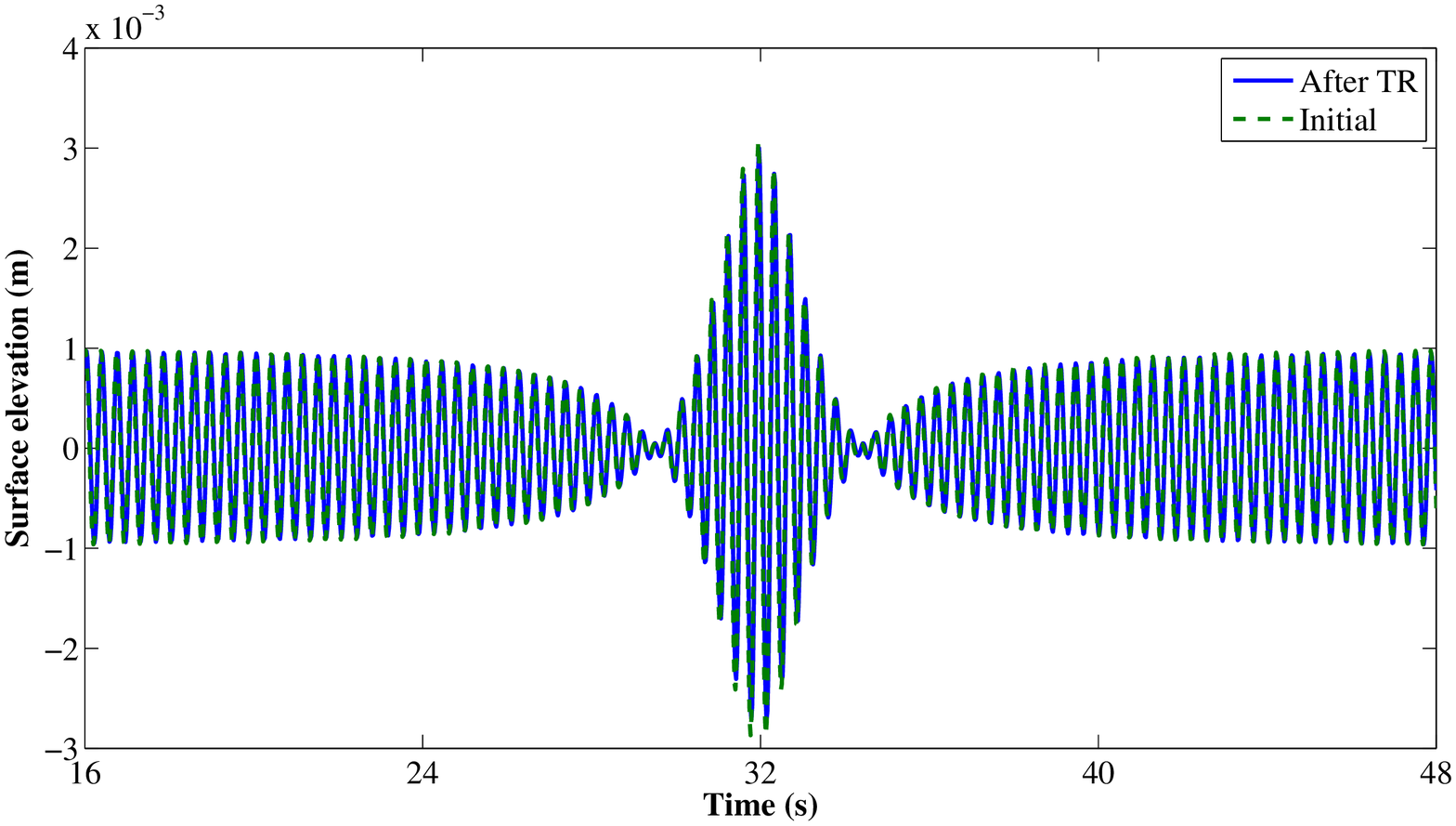}
\includegraphics[width=\linewidth]{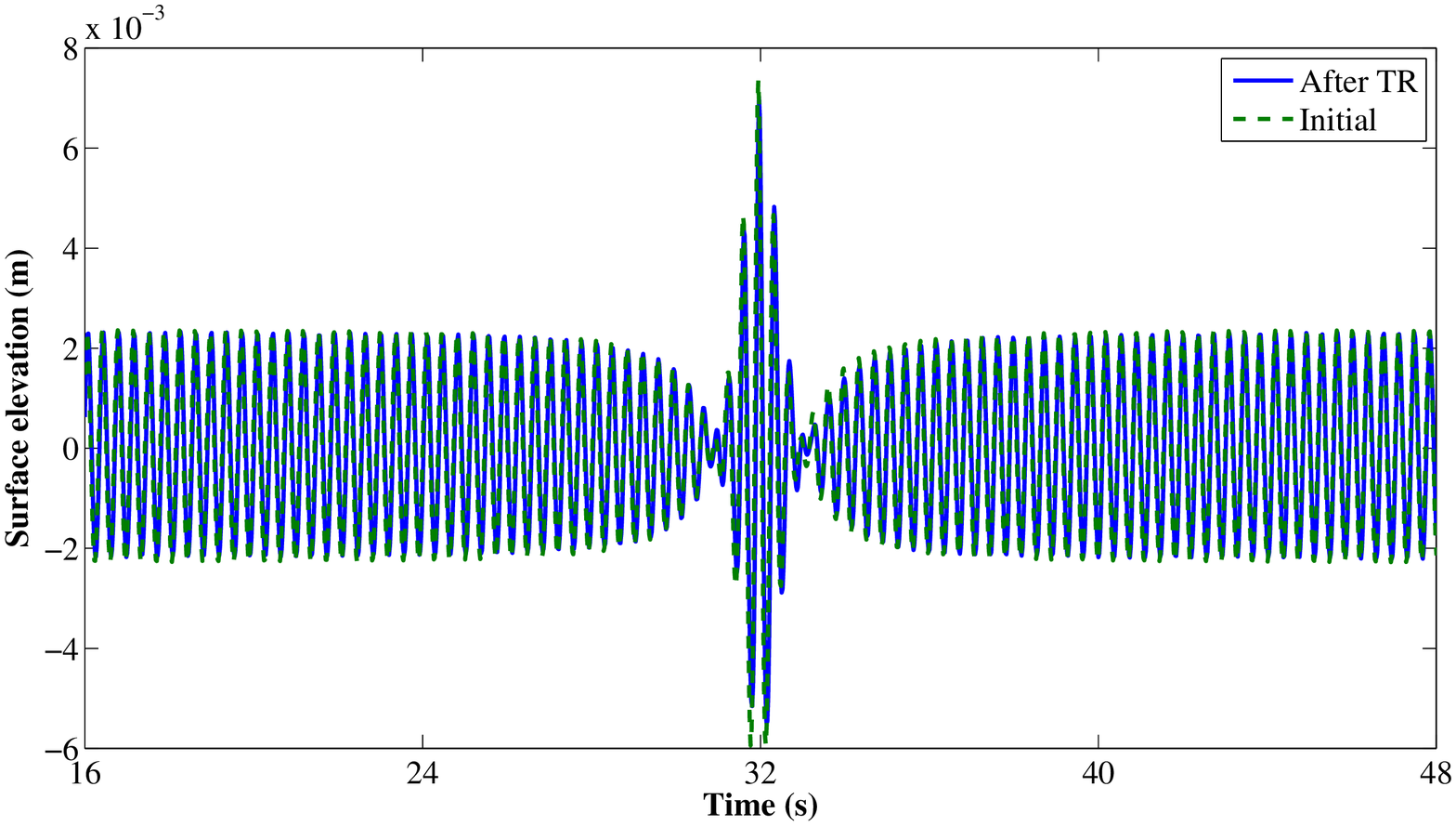}
\includegraphics[width=\linewidth]{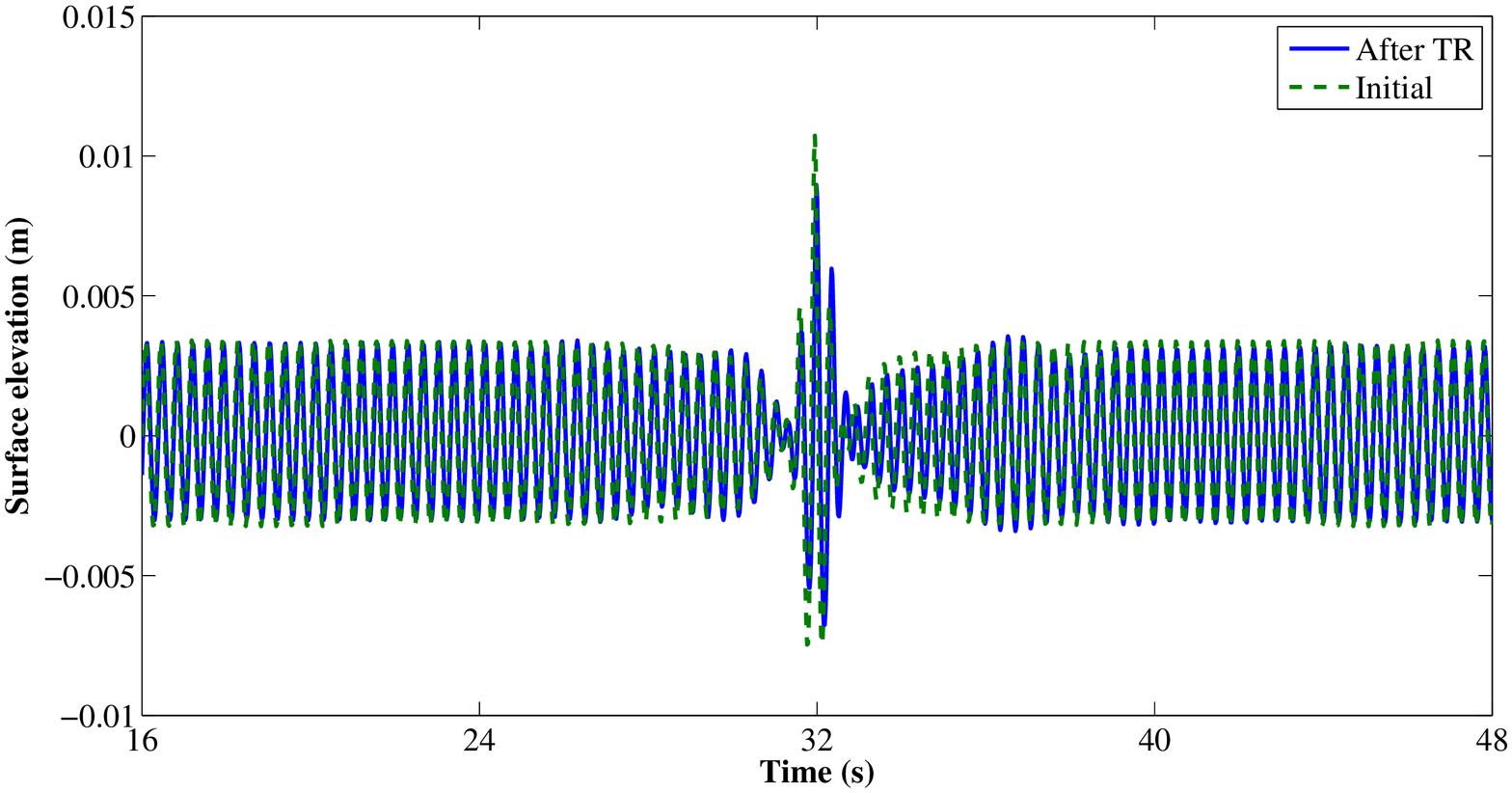}
\includegraphics[width=\linewidth]{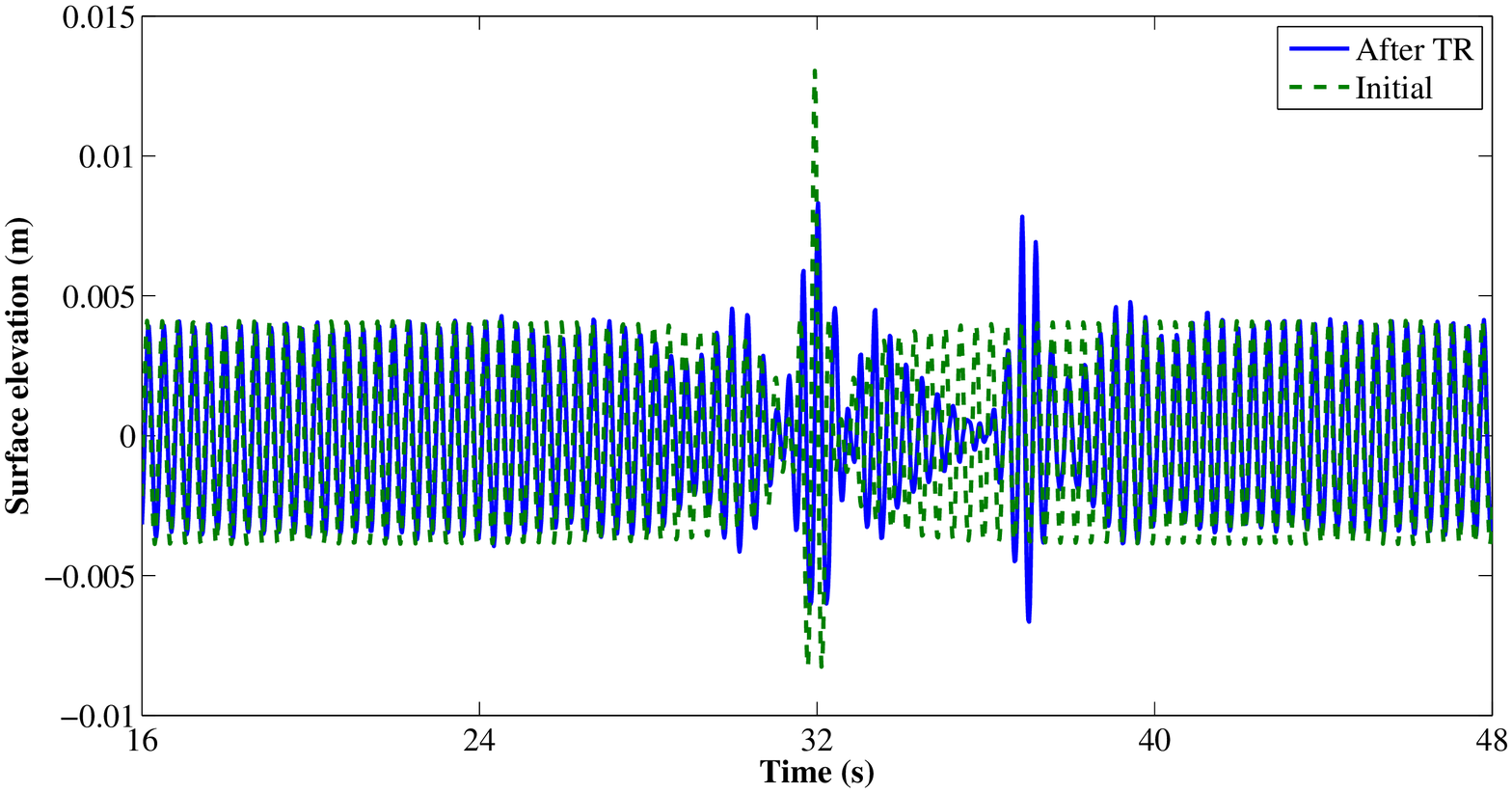}
\caption{Comparison of the Peregrine surface profile at maximal compression, as initially generated with respect to the NLS theory (dashed line) and after TR refocusing (solid line), for $kx_M=270$ and $ka\in\{0.03;0.07;0.1;0.12\}$ (from Top to Bottom).}
\label{fig:Peregrine_TR_kx_270}
\end{center}
\end{figure}

The complexity of the wave pattern at the final stage of demodulation, as a matter of deviation from the constant background, is clearly perceived. An increasing steepness leads as expected to a more complex wave pattern at a fixed location of breather evolution. At the same time, the latter feature is associated to an initial breather modulation which is also quite different when the steepness is changed.  As an example, Figure \ref{fig:Peregrine_TR_kx_270} presents the results obtained after TR refocusing at fixed location of $kx_M=270$, as in the previous case, while varying steepness values, ranging $ka\in\{0.03;0.07;0.10;0.12\}$. 

It can be noticed that looking at a fixed time frame duration, when the steepness is lower, the initial breather appears indeed to be wider. The nonlinearity increases the focusing intensity of the initial localized structure by decreasing the breather's lifetime \cite{chabchoub2012experimental}. The free surface profile with respect to the TR refocusing is almost identical to the analytic solution, when the steepness is lower than $ka\simeq 0.09$. The increase of the steepness is associated to an increase of deviation in terms of phase-shifts. This is similar to what has been observed previously in the dependence study with respect to $kx_M$. The general behavior of the accuracy with respect to this parameter observed in Figure \ref{fig:Peregrine_TR_accuracy_amplitude} is consequently clear.

However, it is important to note that at the largest steepness, the wave field after TR is completely different from the original one. The breather initial structure is completely distorted after the first propagation as seen in Figure \ref{fig:Peregrine_FP_kx_270} and is not refocused during the second propagation after TR. When the considered breather exhibits high nonlinearities, the wave field does not tend to the plane wave solution: successive highly nonlinear groups are created during the propagation. This feature is associated to both high-order dispersive effects as well as high-order nonlinearities. 
These are at play during both propagations and when associated to the chosen wave generation process, they are shown to prevent from time-reversibility and consequently limit the application range of the TR technique.

\begin{figure}[h!tbp]
\begin{center}
\includegraphics[width=\linewidth]{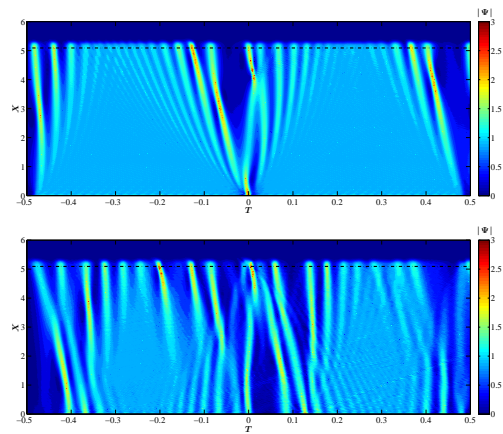}
\caption{Space and time evolution of the Peregrine breather for $ka=0.12$ and $kx_M=480$ (dash-dotted line) in the NWT. (Top) Demodulation (Bottom) TR refocusing.}
\label{fig:Peregrine_FP_and_TR_2dpt_kx_480_ka_0.12}
\end{center}
\end{figure}

Figure \ref{fig:Peregrine_FP_and_TR_2dpt_kx_480_ka_0.12} shows the space-time view of the TR procedure for the case $ka=0.12$ and $kx_M=480$. The corresponding scaled propagating distance $X_M \simeq 5.0$ is depicted on the figure.

Clearly, the TR technique is failing and is not valid in this configuration and it is seen that the breaking of time-reversibility of the phenomena studied appears obvious within these framework. From the beginning of the TR refocusing, noticeable differences can be observed between both propagations, pointing out the importance of the wave generation in the TR procedure.\\

As a conclusion, the refocusing of extreme waves with TR exhibits some limitations. The main parameters in this concern are the nonlinearity of the considered wave field as well as the considered propagating distance. Nevertheless, as seen in Figs. \ref{fig:Peregrine_TR_accuracy_amplitude} \& \ref{fig:Peregrine_2_TR_accuracy_amplitude}, the applicability of process is valid for wide range of realistic parameter choice and is efficient up to high local steepness of the wave pattern and very long propagating distances.




\section{Conclusion} 

To conclude, we reported a detailed numerical study on the applicability of TR mirrors to water waves. A highly nonlinear numerical model based on HOS method has been set-up to reproduce the complete physical configuration of wave generation in a uni-directional NWT: generation of waves by means of a wave maker, absorption of reflecting waves through a beach, etc. The corresponding HOS-NWT has been validated with experimental results presented in \cite{chabchoub2014time}. This numerical set-up enables us to successfully apply the TR method to refocus different analytic solutions of the NLS, namely the stationary envelope soliton and doubly-localized Peregrine-type breathers of first- and second-order. Note that the time-symmetry can be also used to optimize initial conditions for an accurate focusing of water waves \cite{shemer2007evolution}.

The applicability of the TR procedure has been assessed over a wide range of propagating distances $kx_M$ and initial wave steepness $ka$, beyond laboratory limitations. Focusing our attention to pulsating solutions (breathers), the study has demonstrated that the TR procedure will allow an accurate TR refocusing of the wave field at a given propagating distance, when the local steepness of the localized structures does not exceed a given threshold with respect to the propagation distances. We identified that an accurate TR refocusing is achieved for $kx_M\simeq 270$ and a corresponding local steepness up to $ka \simeq 0.27$. Changing the propagating distance will obviously induce a different threshold.

At the same time, the refocusing of extreme waves with respect to TR exhibits some limitations. The time-reversibility of water waves propagation appears to be improper for large steepness and large propagating distances. The practical TR procedure for water waves in the configuration of a wave basin relies on some assumptions resulting in this loss of reversibility. Primarily, this is due to the use of the only free surface elevation for TR associated to the approximation of a linear conversion between free surface elevation and wave maker's motion. These will induce small discrepancies in the generated sea state at the wave maker's location (which are obviously larger with larger nonlinearity). Then, the high-order nonlinearities and dispersive effects associated to the wave propagation will induce possible large discrepancies in the final TR refocusing. The applicability of the method is consequently driven by those two parameters, namely $ka$ and $kx_M$. The present study has shown these latter two parameters are closely linked in describing the limitations of the TR technique with respect to the modulation instability length scales in the case of breathers.


Nevertheless, the applicability of the TR to nonlinear waves for the quantified applicability range is quite remarkable, taking into account the laboratory as well as numerical noise and dissipation effects, always present, when performing numerical and laboratory tests. Furthermore, we emphasize the importance of this study with respect to the robustness of the TR technique to chaotic motions of the water wave field, which is known to arise in the context of modulation instability \cite{caponi1982instability,yasuda1997roles}.

Finally, we emphasize that future interdisciplinary work may be motivated from these results. It is well-known that the NLS describes the propagation of wave in a wide range of nonlinear dispersive media \cite{dudley2009modulation,kibler2010peregrine,onorato2013rogue,dudley2014instabilities}. Furthermore, the reproduction of RWs in wave basins is of significant interest in ocean engineering and has been shown to be very complicated with classical methods \cite{ducrozet2016equivalence}. The idea is to reconstruct oceanic extreme wave, such as the New Year Wave, often used as a dedicated model for RWs in ocean \cite{haver2004possible}. It is also important to address multi-directionality and irregular wave conditions in order to accurately assert the validity and limitation of the TR in these conditions. Work with respect to these latter aspects has started. 

\bibliography{myref}

\end{document}